\documentclass[aps,prb,showpacs,twocolumn,superscriptaddress]{revtex4-1}

\usepackage{graphicx,amsmath,amssymb}
\usepackage[usenames]{color}
\usepackage{float}
\usepackage[T1]{fontenc}

\usepackage[colorlinks=true, citecolor=blue, urlcolor=blue, linkcolor=blue ]{hyperref}
\hypersetup{breaklinks=true}
\begin{document}

\bibliographystyle{apsrev4-1}

\title{Friedel oscillation in non-Fermi liquid: Lesson from exactly solvable Hatsugai-Kohmoto model}

\author{Miaomiao Zhao}
\affiliation{School of Physical Science and Technology $\&$ Key Laboratory for
Magnetism and Magnetic Materials of the MoE, Lanzhou University, Lanzhou 730000, China}
\affiliation{Lanzhou Center for Theoretical Physics, Key Laboratory of Theoretical Physics of Gansu Province}

\author{Wei-Wei Yang}
\affiliation{School of Physical Science and Technology $\&$ Key Laboratory for
Magnetism and Magnetic Materials of the MoE, Lanzhou University, Lanzhou 730000, China}
\affiliation{Lanzhou Center for Theoretical Physics, Key Laboratory of Theoretical Physics of Gansu Province}

\author{Hong-Gang Luo}
\affiliation{School of Physical Science and Technology $\&$ Key Laboratory for
Magnetism and Magnetic Materials of the MoE, Lanzhou University, Lanzhou 730000, China}
\affiliation{Lanzhou Center for Theoretical Physics, Key Laboratory of Theoretical Physics of Gansu Province}
\affiliation{Beijing Computational Science Research Center, Beijing 100084, China}

\author{Yin Zhong}
\email{zhongy@lzu.edu.cn}
\affiliation{School of Physical Science and Technology $\&$ Key Laboratory for
Magnetism and Magnetic Materials of the MoE, Lanzhou University, Lanzhou 730000, China}
\affiliation{Lanzhou Center for Theoretical Physics, Key Laboratory of Theoretical Physics of Gansu Province}

\begin{abstract}
When non-magnetic impurity immerses in Fermi sea, a regular modulation of charge density around impurity will appear and such phenomena is called Friedel oscillation (FO). Although both Luttinger liquid and Landau Fermi liquid show such characteristic oscillation, FO in generic non-Fermi liquid (NFL) phase is still largely unknown. Here, we show that FO indeed exists in NFL state of an exactly solvable model, i.e. the Hatsugai-Kohmoto model which has been intensively explored in recent years. Combining T-matrix approximation and linear-response-theory, an interesting picture emerges, if two interaction-induced quasi-particles bands in NFL are partially occupied, FO in this situation is determined by a novel structure in momentum space, i.e. the 'average Fermi surface' (average over two quasi-particle Fermi surface), which highlights the inter-band particle-hole excitation. We hope our study here provides a counterintuitive example in which FO with Fermi surface coexists with NFL quasi-particle, and it may be useful to detect hidden 'average Fermi surface' structure in other correlated electron systems.
\end{abstract}

\maketitle
\section{Introduction}
Recently, an exactly solvable many-body fermionic model with an infinite-range interaction, i.e. the Hatsugai-Kohmoto (HK) model,\cite{Hatsugai1992,Baskaran1991,Hatsugai1996} has been hotly studied.\cite{Phillips2018,Yeo2019,Phillips2020,Yang2021,Zhu2021,Zhao2022,Setty2021,Mai2022,Huang2022,Li2022,Setty2020,Setty2021b,Zhong2022,Wang2023,Zhong2023,Leeb2023,Wang2023b,Setty2023}
In contrast with celebrated and more familiar Sachdev-Ye-Kitaev, Kitaev's toric code and honeycomb model with either quenched disorder or local $Z_{2}$ gauge symmetry,\cite{Sachdev,Maldacena,Chowdhury,Kitaev1,Kitaev2,Prosko,Zhong2013,Smith2017} the original HK model has translation invariance with topologically trivial nature, but surprisingly, it provides a strictly exact playground for non-Fermi liquid (NFL) and featureless Mott insulator in any spatial dimension, which is rare in statistical mechanics and condensed matter physics.

The solvability of HK model results from its locality in momentum space and one can diagonalize HK Hamiltonian (just diagonal $4\times4$-matrix) for each momentum. The current studies have mainly focused on an interesting extension of HK model, i.e. the superconducting instability from the intrinsic NFL state in HK model,\cite{Phillips2020} which is inspired by ubiquitous NFL behaviors and their link to unconventional superconductivity in cuprate, iron-based superconductors (SC) and many heavy fermion compounds. Unexpected properties such as topological $s$-wave pairing and two-stage superconductivity have been discovered.\cite{Zhu2021,Zhao2022,Li2022} However, before comparing these novel theoretical predictions with real-world unconventional SC in cuprate or heavy fermion systems, one should note that non-magnetic impurity is essential to explain realistic thermodynamic and transport date in SC, e.g. impurity effect changes linear-$T$ behavior in superfluid density of the nodal $d$-wave paring state into $T^{2}$ form.\cite{Xiang2022} However, to our knowledge, the mentioned non-magnetic impurity effect has not been investigated in HK model, let alone the superconducting HK system.

For metals, it is well-known that non-magnetic impurity immersed in Fermi sea induces a regular modulation of charge density around impurity, e.g. the Friedel oscillation (FO).\cite{Friedel1952} When involving electron-electron interaction, both the Luttinger liquid in one spatial dimension ($d=1$) and the Landau Fermi liquid (FL) show such characteristic oscillation.\cite{Egger1995,Leclair1996,Simion2005,Chatterjee2015,White2002,Soffing2009} The origin of FO is generally believed to tie to the $2k_{F}$ singularity of density-density correlation in the system with sharp Fermi surface, thus even quantum spin liquid with ghost (spinon) Fermi surface may show signature of FO.\cite{Mross2011,Zhou} Consequently, FO can act as diagnosis for fermionic system with well-defined Fermi surface whatever its FL or NFL nature and may shed light on how to detect putative NFL state in realistic quantum materials proposed by existing effective field theory or slave-particle theory.\cite{Lee2005,Senthil2008a,Senthil2008b,Metlitski2010a,Metlitski2010b,Mross2010,Senthil2004,Paul2007,Zhong2012,Wolfle2012,Abrahams2014}

Therefore, considering the need of exploration on FO for generic NFL phases and the request from superconducting HK models, in this work, we take a first step study on this timely issue. Specifically, we focus on the simplest but essential case in non-magnetic impurity effect, i.e. the possible FO in single impurity problem.

To our surprise, we find that the conventional wisdom of FO must be extended, since the NFL phase in HK model with two Fermi surface but no FL quasi-particle, indeed shows clear signature of FO. As a matter of fact, FO in this situation is determined by a novel structure in momentum space, i.e. the 'average Fermi surface', this means the average over the mentioned two Fermi surface and it results from the inter-band particle-hole excitation.

After all, our study here provides a counterintuitive example in which Fermi surface coexisting with NFL quasi-particle can support the existence of FO, and we expect that it may be interesting to detect hidden 'average Fermi surface' structure in other correlated electron systems.
\begin{figure}
\includegraphics[width=0.75\linewidth]{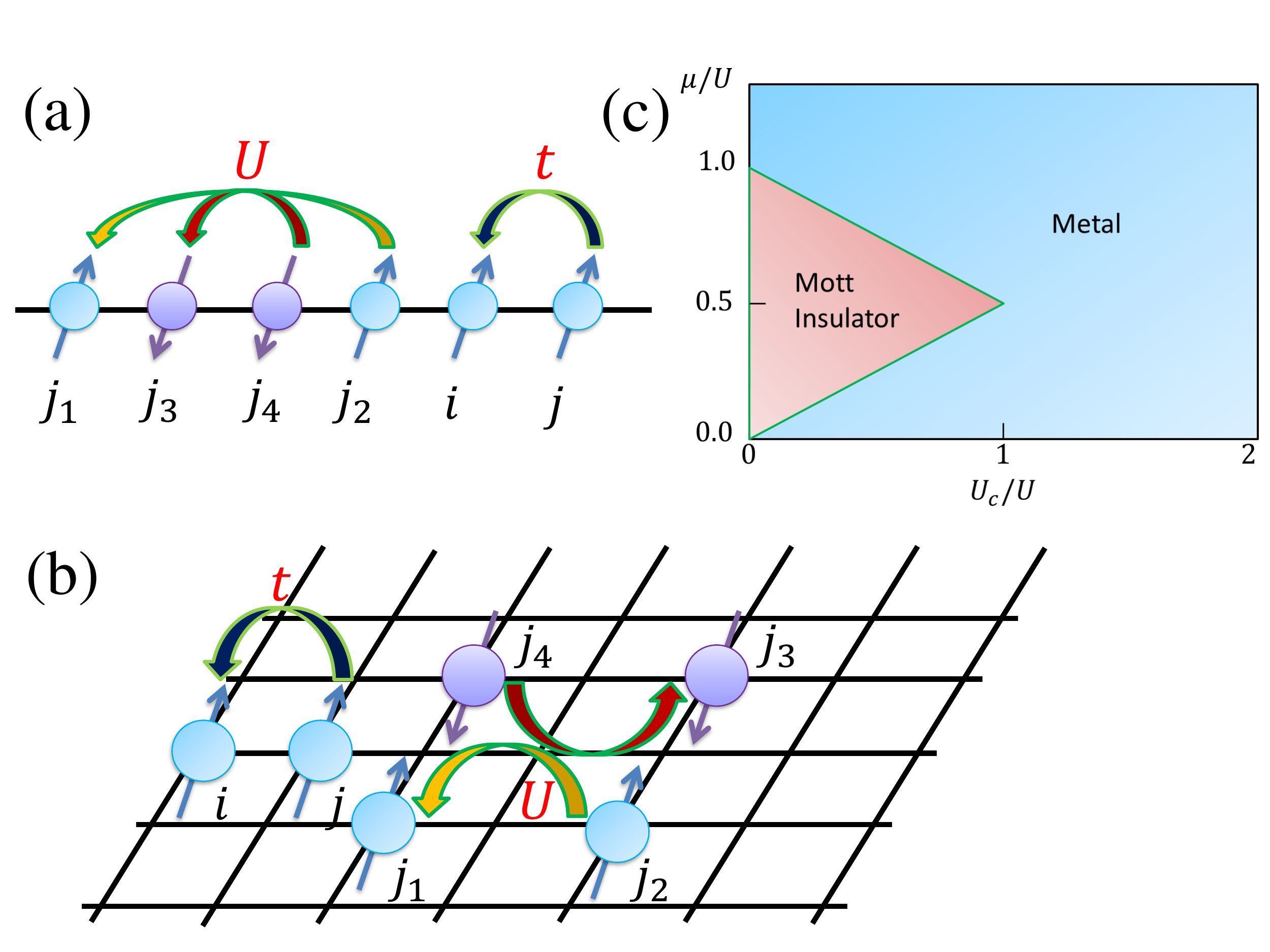}
\caption{\label{fig:fig_0} (a) The Hatsugai-Kohmoto (HK) model in one spatial dimension and (b) on a square lattice with hopping $t$ and interaction $U$. (c) The exact ground-state phase diagram for HK model exhibits a
Mott insulator and a non-Fermi-liquid-like metal. ($\mu$ denotes chemical potential, $U_{c}=W$ and $W$ is band-width) The transition from metallic state to
gapped Mott insulating phase belongs to the universality of the continuous Lifshitz transition.}
\end{figure}

The remaining part of this article is organized as follows. In Sec.~\ref{sec1}, we give a quick review of HK model, which will be useful in next sections. Sec.~\ref{sec2} is devoted to the discussion of FO in terms of $T$-matrix approximation and the linear-response theory. Discussion will be given in Sec.~\ref{sec3}. Finally, a brief summary is given in Sec.~\ref{sec4}.

\section{The Hatsugai-Kohmoto model}\label{sec1}
The HK model we study has the following form, (see also Fig.~\ref{fig:fig_0}(a) and (b) for illustration of HK model in one spatial dimension and on a square lattice)
\begin{eqnarray}
\hat{H}&=&-\sum_{i,j,\sigma}t_{ij}\hat{c}_{i\sigma}^{\dag}\hat{c}_{j\sigma}-\mu\sum_{j\sigma}\hat{c}_{j\sigma}^{\dag}\hat{c}_{j\sigma}\nonumber\\
&+&\frac{U}{N_{s}}\sum_{j_{1},j_{2},j_{3},j_{4}}\delta_{j_{1}+j_{3}=j_{2}+j_{4}}
\hat{c}_{j_{1}\uparrow}^{\dag}\hat{c}_{j_{2}\uparrow}\hat{c}_{j_{3}\downarrow}^{\dag}\hat{c}_{j_{4}\downarrow}\label{eq1}.
\end{eqnarray}
Here, $\hat{c}_{j\sigma}^{\dag}$ is the creation operator of conduction electron (called $c$-electron for simplicity) at site $j$ with spin $\sigma=\uparrow,\downarrow$ and it satisfies anti-commutative relation $[\hat{c}_{i\sigma},\hat{c}_{j\sigma'}^{\dag}]_{+}=\delta_{i,j}\delta_{\sigma,\sigma'}$. $t_{ij}$ are hopping integral between $i,j$ sites. Furthermore, chemical potential $\mu$ has been added to fix electron's density. $N_{s}$ is the number of sites. The last term of $\hat{H}$ is the HK interaction,\cite{Hatsugai1992} unlike the usual on-site Hubbard interaction $U\sum_{j}\hat{c}_{j\uparrow}^{\dag}\hat{c}_{j\uparrow}\hat{c}_{j\downarrow}^{\dag}\hat{c}_{j\downarrow}$, HK interaction is an infinite-range interaction between four electrons but preserves the center of motion due to the constraint of Dirac's $\delta$ function. This interaction plays a fundamental role in solving this model as we will see later. Amusingly, one may note that the HK interaction indeed includes the Hubbard interaction if we consider a two-site version of HK model. However, the true effect of the latter one for HK-like models beyond perturbative treatment is still unknown and such issue seems to be important for our further understanding on HK-like systems.

Importantly, Eq.~\ref{eq1} is local in momentum space after Fourier transformation (i.e. $\hat{c}_{j\sigma}=\frac{1}{\sqrt{N_{s}}}\sum_{k}e^{ikR_{j}}\hat{c}_{k\sigma}$) and the resultant Hamiltonian reads as $\hat{H}=\sum_{k}\hat{H}_{k}$,
\begin{eqnarray}
\hat{H}_{k}&=&\sum_{\sigma}(\varepsilon_{k}-\mu)\hat{c}_{k\sigma}^{\dag}\hat{c}_{k\sigma}+U
\hat{c}_{k\uparrow}^{\dag}\hat{c}_{k\uparrow}\hat{c}_{k\downarrow}^{\dag}\hat{c}_{k\downarrow}\label{eq2},
\end{eqnarray}
where $\varepsilon_{k}$ are dispersion of electrons. It is emphasized that the locality of above Hamiltonian stems from infinite-range HK interaction preserving center of motion. In contrast, the Hubbard interaction in momentum space is rather nonlocal as $U\sum_{k,k',q}\hat{c}_{k+q\uparrow}^{\dag}\hat{c}_{k\uparrow}\hat{c}_{k'-q\downarrow}^{\dag}\hat{c}_{k'\downarrow}$, thus it cannot lead to solvability for $d>1$.

Now, if we choose Fock state
\begin{equation}
|n_{1},n_{2}\rangle\equiv
(\hat{c}_{k\uparrow}^{\dag})^{n_{1}}|0\rangle(\hat{c}_{k\downarrow}^{\dag})^{n_{2}}|0\rangle\label{eq3}
\end{equation}
with $n_{i}=0,1$ as basis, $\hat{H}_{k}$ can be written as a diagonal $4\times4$ matrix, whose eigen-energy is $0,\varepsilon_{k}-\mu,\varepsilon_{k}-\mu,2(\varepsilon_{k}-\mu)+U$ and the corresponding eigen-state is $|0\rangle_{k}\equiv|00\rangle,|\sigma=\uparrow\rangle_{k}\equiv|10\rangle,|\sigma=\downarrow\rangle_{k}\equiv|01\rangle,|\uparrow\downarrow\rangle_{k}\equiv|11\rangle$, which means states are empty, single occupied with spin-up and spin-down, and double occupied.

Therefore, the many-body ground-state of $\hat{H}$ is just the direct-product state of each $\hat{H}_{k}$'s ground-state, i.e. $|\Psi_{g}\rangle=\prod_{k\in\Omega_{0}}|0\rangle_{k}\prod_{k\in\Omega_{1}}|\sigma\rangle_{k}\prod_{k\in\Omega_{2}}|\uparrow\downarrow\rangle_{k}$. ($\Omega_{0},\Omega_{1},\Omega_{2}$ are the momentum range for different occupation) Because states with spin-up or down electron in $\Omega_{1}$ is degenerated without external magnetic field, the ground-state of HK model has huge degeneracy. This point must be kept in mind if one performs numerical calculation like exact diagonalization (ED) which only selects one of ground-states.\cite{Zhong2022}

If $\Omega_{0}=\Omega_{2}=0$, the system is a Mott insulator which happens when $U>U_{c}=W$ with $W$ being the bandwidth of $c$-electron. Otherwise, we obtain a metallic state with NFL properties, e.g. violation of Luttinger theorem, Haldane's exclusion
statistics and Curie-like spin susceptibility.\cite{Phillips2020,Hatsugai1996,Vitoriano2000} It is interesting to note that as a result of HK interaction which preserves the center of motion, NFL states do not have collective mode in charge degree of freedom, such as the plasmon in Coulomb electron gas or zero sound in FL.\cite{Pines1966,Ketterson2016} Moreover, the transition from metallic state to
gapped Mott insulating phase belongs to the universality of the continuous Lifshitz transition, in which the chemical potential-tuning and the interaction-tuning Mott transition have identical critical exponents.\cite{Vitoriano2000,Continentino}(see also Fig.~\ref{fig:fig_0}(c)) Similarly, excited states and their energy are easy to be constructed, so $\hat{H}$ (Eq.~\ref{eq1}) has been solved since all eigen-states and eigen-energy are found.

For our purpose, it is useful to present the single-particle Green's function and some ground-state or thermodynamic quantities for HK model. For example, the single-particle Green's function can be obtained in terms of equation of motion,\cite{Hubbard1963}(see Appendix.~\ref{ap_A}) which is read as
\begin{eqnarray}
G_{\sigma}(k,\omega)&=&\frac{1+\frac{U\langle\hat{n}_{k\bar{\sigma}}\rangle}{\omega-(\varepsilon_{k}-\mu+U)}}{\omega-(\varepsilon_{k}-\mu)}\nonumber\\
&=&\frac{1-\langle\hat{n}_{k\bar{\sigma}}\rangle}{\omega-(\varepsilon_{k}-\mu)}+\frac{\langle\hat{n}_{k\bar{\sigma}}\rangle}{\omega-(\varepsilon_{k}-\mu+U)}\label{eq4}
\end{eqnarray}
where $\langle\hat{n}_{k\bar{\sigma}}\rangle$ is the expectation value of electron number operator $\hat{n}_{k\bar{\sigma}}=\hat{c}^{\dag}_{k\bar{\sigma}}\hat{c}_{k\bar{\sigma}}$ with spin $\bar{\sigma}=-\sigma$. The pole structure implies that there exist two quasi-particle bands as
\begin{equation}
E_{k-}=\varepsilon_{k}-\mu,~~~~E_{k+}=\varepsilon_{k}-\mu+U,\label{eq5}
\end{equation}
which corresponds to holon $\hat{h}_{k\sigma}=\hat{c}_{k\sigma}(1-\hat{n}_{k\bar{\sigma}})$ and doublon $\hat{d}_{k\sigma}=\hat{c}_{k\sigma}\hat{n}_{k\bar{\sigma}}$. In fact, the related Green's function is found to be
\begin{equation}
\langle\langle \hat{h}_{k\sigma}|\hat{h}_{k\sigma}^{\dag}\rangle\rangle_{\omega}=\frac{1-\langle \hat{n}_{k\bar{\sigma}}\rangle}{\omega-\varepsilon_{k}+\mu}=\frac{1-\langle \hat{n}_{k\bar{\sigma}}\rangle}{\omega-E_{k-}},\nonumber
\end{equation}
\begin{equation}
\langle\langle \hat{d}_{k\sigma}|\hat{d}_{k\sigma}^{\dag}\rangle\rangle_{\omega}=\frac{\langle \hat{n}_{k\bar{\sigma}}\rangle}{\omega-\varepsilon_{k}+\mu-U}=\frac{\langle \hat{n}_{k\bar{\sigma}}\rangle}{\omega-E_{k+}}.\nonumber
\end{equation}
Thus, we see that the elementary excitations of HK model are holon and doublon. But we should emphasize that holon or doublon does not satisfy standard fermionic anti-commutative relation and cannot adiabatically evolve into $U=0$ limit, thus they are not FL-like quasi-particle.

Since we are interested in paramagnetic states, we have $n_{k}=\langle\hat{n}_{k\sigma}\rangle=\langle\hat{n}_{k\bar{\sigma}}\rangle$ and it is straightforward to find
\begin{equation}
  n_{k}=\frac{f_{F}(E_{k-})}{f_{F}(E_{k-})+1-f_{F}(E_{k+})}\label{eq6}
\end{equation}
with the help of spectral theorem of $G_{\sigma}(k,\omega)$. ($f_{F}(x)=1/(e^{x/T}+1)$ is the Fermi distribution function)

Next, at finite-$T$, the thermodynamics of HK model is determined by
its free energy density $f$, which is related to partition function $\mathcal{Z}$ as
\begin{eqnarray}
f=-\frac{T}{N_{s}}\ln\mathcal{Z},~~\mathcal{Z}=\mathrm{Tr }e^{-\beta \hat{H}}=\prod_{k}\mathrm{Tr} e^{-\beta \hat{H}_{k}}=\prod_{k}f_{k}.\nonumber
\end{eqnarray}
Here, one notes that the partition function is easy to calculate since each $k$-state contributes independently. We have defined $f_{k}=1+2z_{k}+z_{k}^{2}e^{-\beta U}$ and $z_{k}=e^{-\beta(\varepsilon_{k}-\mu)}$. Then, the typical thermodynamic quantity, i.e. the heat capacity, is calculated by standard thermodynamic relation $C_{V}=-T\frac{\partial^{2}f}{\partial T^{2}}$. In addition to $C_{V}$, one can also calculate spin susceptibility $\chi_{s}$ if one inserts Zeeman energy term $\hat{H}_{h}=-B(\hat{c}_{k\uparrow}^{\dag}\hat{c}_{k\uparrow}-\hat{c}_{k\downarrow}^{\dag}\hat{c}_{k\downarrow})$ into Hamiltonian $\hat{H}_{k}$. Then, it follows that the magnetization $M=-\frac{\partial f}{\partial B}$ and $\chi_{s}=\frac{\partial M}{\partial B}=-\frac{\partial^{2}f}{\partial B^{2}}$.

At zero temperature, the free energy density reduces into the ground-state energy density, which has very simple expression,
\begin{equation}
e_{g}=\frac{1}{N_{s}}\sum_{k}[E_{k-}\theta(-E_{k-})+E_{k+}\theta(-E_{k+})],\nonumber
\end{equation}
where $\theta(x)$ is the standard unit-step function ($\theta(x)=1$ for $x>0$ and $\theta(x)=0$ if $x<0$). Therefore, the electron density at $T=0$ is found to be
\begin{equation}
n=-\frac{\partial e_{g}}{\partial \mu}=\frac{1}{N_{s}}\sum_{k}[\theta(-E_{k-})+\theta(-E_{k+})].\label{eq7}
\end{equation}
\section{Single impurity in HK model}\label{sec2}
In this section, we study the impurity effect in HK model. It is well-known that for non-interacting Fermi gas and interacting FL or Luttinger liquid, the electron density around impurity shows characteristic oscillation called FO.
\subsection{$T$-matrix approximation}
Now, we consider the effect of a single impurity, which is assumed to locate on zero-th site and only electron on this site feels its scattering, thus we have the following impurity Hamiltonian:
\begin{equation}
\hat{H}_{imp}=V\sum_{\sigma}\hat{c}_{0\sigma}^{\dag}\hat{c}_{0\sigma}=\frac{V}{N_{s}}\sum_{k,k',\sigma}\hat{c}_{k\sigma}^{\dag}\hat{c}_{k'\sigma}\nonumber
\end{equation}
Here, $V$ is the strength of impurity potential and the second term of the right-hand side is the Hamiltonian in momentum space.

If HK interaction is turned off, one can solve the non-interacting electron problem $\hat{H}+\hat{H}_{imp}$ in terms of $T$-matrix formalism, which means the Green's function satisfies the following equations,\cite{Coleman2015}
\begin{eqnarray}
&&G_{\sigma}^{(0)}(k,k',\omega)=\delta_{k,k'}G_{\sigma}^{(0)}(k,\omega)+G_{\sigma}^{(0)}(k,\omega)T_{\sigma}(\omega)G_{\sigma}^{(0)}(k',\omega)\nonumber\\
&&T_{\sigma}(\omega)=\frac{V/N_{s}}{1-VF_{\sigma}(\omega)},~~F_{\sigma}(\omega)=\frac{1}{N_{s}}\sum_{k}G_{\sigma}^{(0)}(k,\omega)\nonumber
\end{eqnarray}
Here, we have defined the so-called $T$-matrix $T_{\sigma}(\omega)$, which encodes the effect of impurity scattering. Meanwhile, $G^{(0)}$ denotes the Green's function for $U=0$ and
\begin{eqnarray}
G_{\sigma}^{(0)}(k,\omega)=\frac{1}{\omega-(\varepsilon_{k}-\mu)}.\nonumber
\end{eqnarray}

However, one can see that when $U\neq0$, $\hat{H}_{imp}$ mixes different momentum sectors of original HK model (Eq.\ref{eq1}), therefore, the solvability of HK model is lost and we can only obtain accurate results from numerical computation like ED.

To proceed, let us use the above $T$-matrix formalism as an approximation and we expect that such approximated treatment may be appropriate if impurity strength $V$ is not large. Then, we replace non-interacting Green's function $G^{(0)}$ with interacting $G$ without impurity (Eq.~\ref{eq4}) as what has usually been done in dynamic mean-field theory study or cuprate superconductor.\cite{Lederer2008,Nowadnick2012,Chatterjee2019} So, we find
\begin{eqnarray}
&&G_{\sigma}(k,k',\omega)\simeq\delta_{k,k'}G_{\sigma}(k,\omega)+G_{\sigma}(k,\omega)T_{\sigma}(\omega)G_{\sigma}(k',\omega)\nonumber\\
&&T_{\sigma}(\omega)=\frac{V/N_{s}}{1-VF_{\sigma}(\omega)},~~F_{\sigma}(\omega)=\frac{1}{N_{s}}\sum_{k}G_{\sigma}(k,\omega).\label{eq8}
\end{eqnarray}

In reality, the effect of single impurity can be observed via the well-known FO,\cite{Friedel1952} which states that the electron's density around impurity shows a characteristic oscillation behavior. For systems with well-defined Fermi surface, such as Landau FL in $d=2,3$ and Luttinger liquid in $d=1$,\cite{Simion2005,Chatterjee2015,Egger1995,Leclair1996,White2002,Soffing2009} the FO behaves as
\begin{equation}
\delta n_{i}\equiv n_{i}-n\sim \frac{\cos(2k_{F}|R_{i}|)}{|R_{i}|^{g}},~~~~|R_{i}|>>1\nonumber
\end{equation}
where $n_{i}$ is the electron's density at $i$-site, $n$ denotes the average electron density without impurity, $R_{i}$ is the distance versus impurity (assumed on $0$-site in our model), $k_{F}$ is Fermi wavevector and $g$ is equal to the spatial dimension for FL,\cite{Simion2005} or determined by interaction strength in Luttinger liquid.\cite{Egger1995,White2002}

Our aim of this section is to examine whether the above FO survives in metallic NFL state of HK model. Mathematically, we can write $\delta n_{i}$ as
\begin{eqnarray}
 \delta n_{i}&=&\frac{1}{N_{s}}\sum_{k,k',\sigma}e^{i(k-k')R_{i}}\langle c_{k'\sigma}^{\dag}c_{k\sigma}\rangle-n\nonumber\\
 &=&\frac{1}{N_{s}}\sum_{k,k',\sigma}e^{i(k-k')R_{i}}\int d\omega f_{F}(\omega)\frac{-1}{\pi}\mathrm{Im}G(k,k',\omega)-n\nonumber\\
 &=&\frac{1}{N_{s}}\sum_{k,k',\sigma}e^{i(k-k')R_{i}}\int d\omega f_{F}(\omega)\frac{-1}{\pi}\mathrm{Im}\delta G_{\sigma}(k,k',\omega).\nonumber
\end{eqnarray}
Here, we have defined the scattering shifted Green's function $\delta G_{\sigma}(k,k',\omega)\equiv G_{\sigma}(k,\omega)T_{\sigma}(\omega)G_{\sigma}(k',\omega)$ and $f_{F}(x)=1/(e^{x/T}+1)$ is the Fermi distribution function. Next, sum over momentum, one finds
\begin{eqnarray}
 &&\delta n_{i}=\sum_{\sigma}\int d\omega f_{F}(\omega)\frac{-1}{\pi}\mathrm{Im}
 \left[\frac{G_{\sigma}(R_{i},\omega)VG_{\sigma}(-R_{i},\omega)}{1-VF_{\sigma}(\omega)}\right]\nonumber\\
 &&\label{eq_del_n}
\end{eqnarray}
Here, $G_{\sigma}(R_{i},\omega)=\frac{1}{N_{s}}\sum_{k}e^{ikR_{i}}G_{\sigma}(k,\omega)$ is the local Green's function on $i$-site. Then, we are able to utilize Eq.~\ref{eq_del_n} to calculate $\delta n_{i}$, such that the check on FO is straightforward.

\begin{figure}
\includegraphics[width=0.9\linewidth]{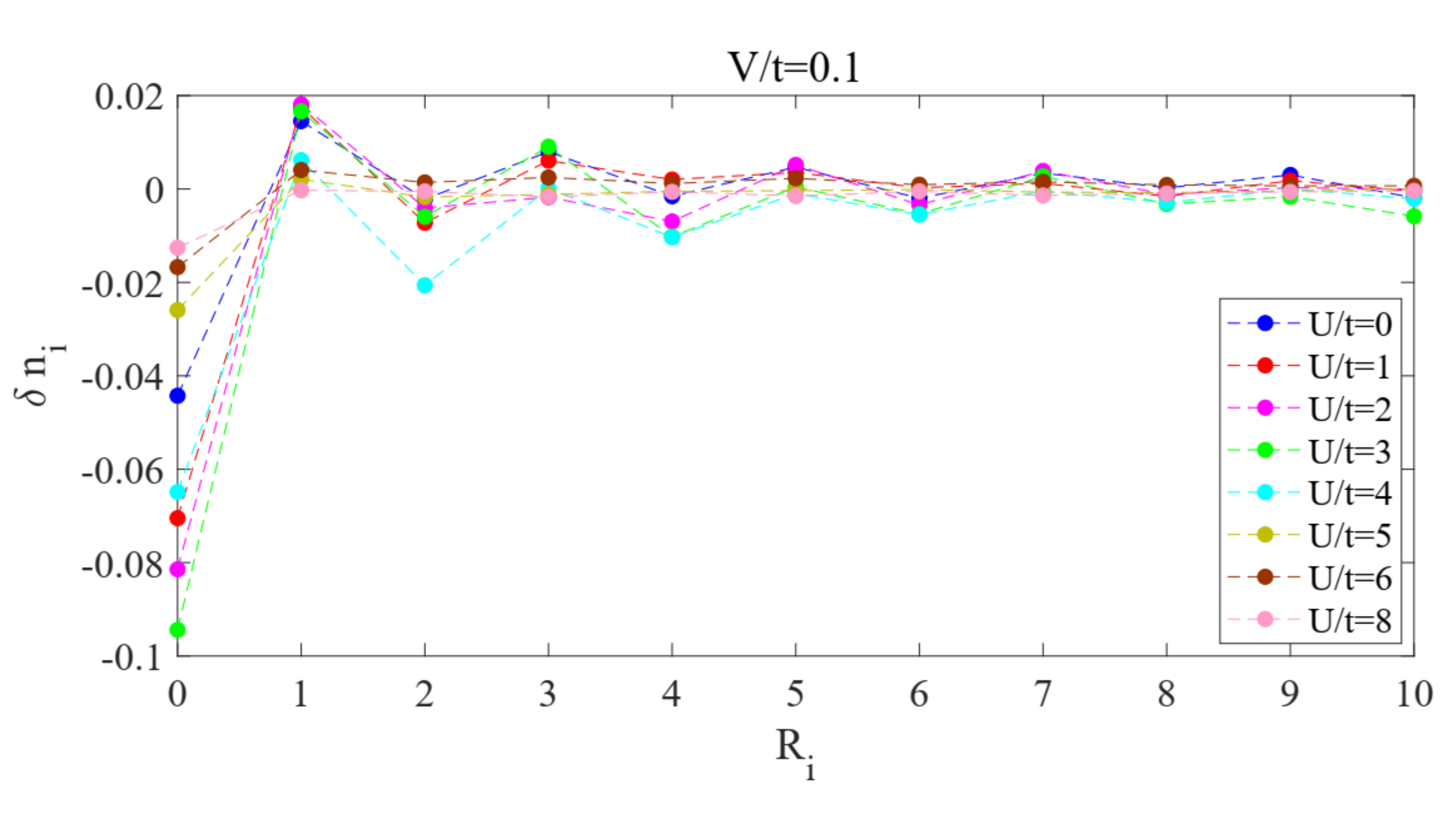}
\caption{\label{fig:V_01} $\delta n_{i}$ for $U/t=0,1,2,3,4,5,6,8$ with $V/t=0.1$ and $\mu=U/2$.}
\end{figure}

\begin{figure}
\includegraphics[width=0.9\linewidth]{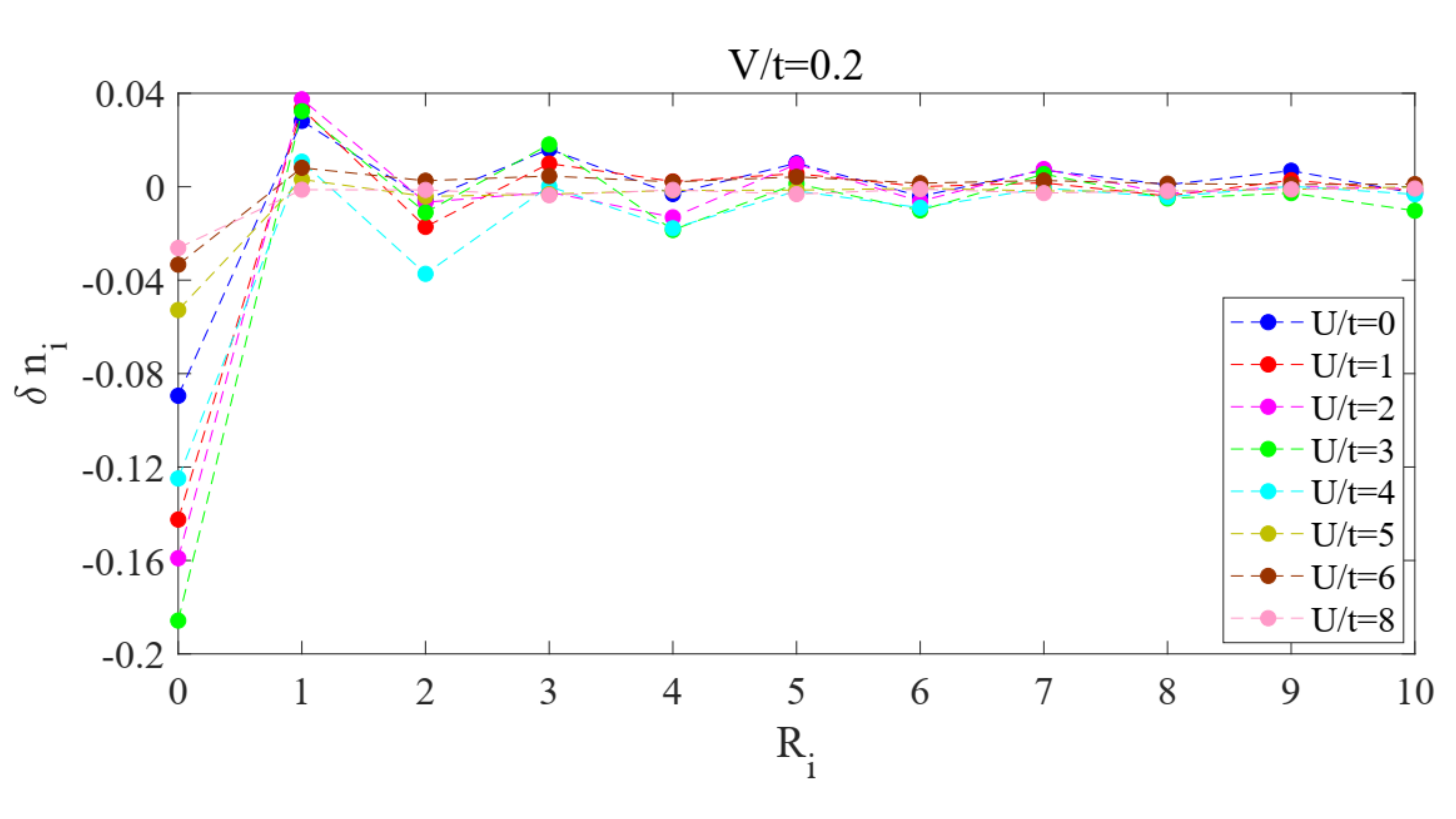}
\caption{\label{fig:V_02} $\delta n_{i}$ for $U/t=0,1,2,3,4,5,6,8$ with $V/t=0.2$ and $\mu=U/2$.}
\end{figure}

In Fig.~\ref{fig:V_01} and \ref{fig:V_02}, we have plotted $\delta n_{i}$ for $V/t=0.1,0.2$ with different $U/t=0,1,2,3,4,5,6,8$ and in $1D$ case. ($\varepsilon_{k}=-2t\cos(k)$) To explore both metallic and insulating phases, we have fixed $\mu=U/2$. Here, we see that if the system is located in metallic phase ($U/t\leq4$), FO is clearly visible since all data is similar to non-interacting case ($U=0$), which shows FO-like behavior as $\delta n_{i}\sim \frac{\cos(2k_{F}|R_{i}|)}{|R_{i}|}$. ($k_{F}=\pi/2$) In contrast, when the ground-state is in the Mott insulating phase ($U/t>4$), signal of FO vanishes.

\begin{figure}
\includegraphics[width=1.0\linewidth]{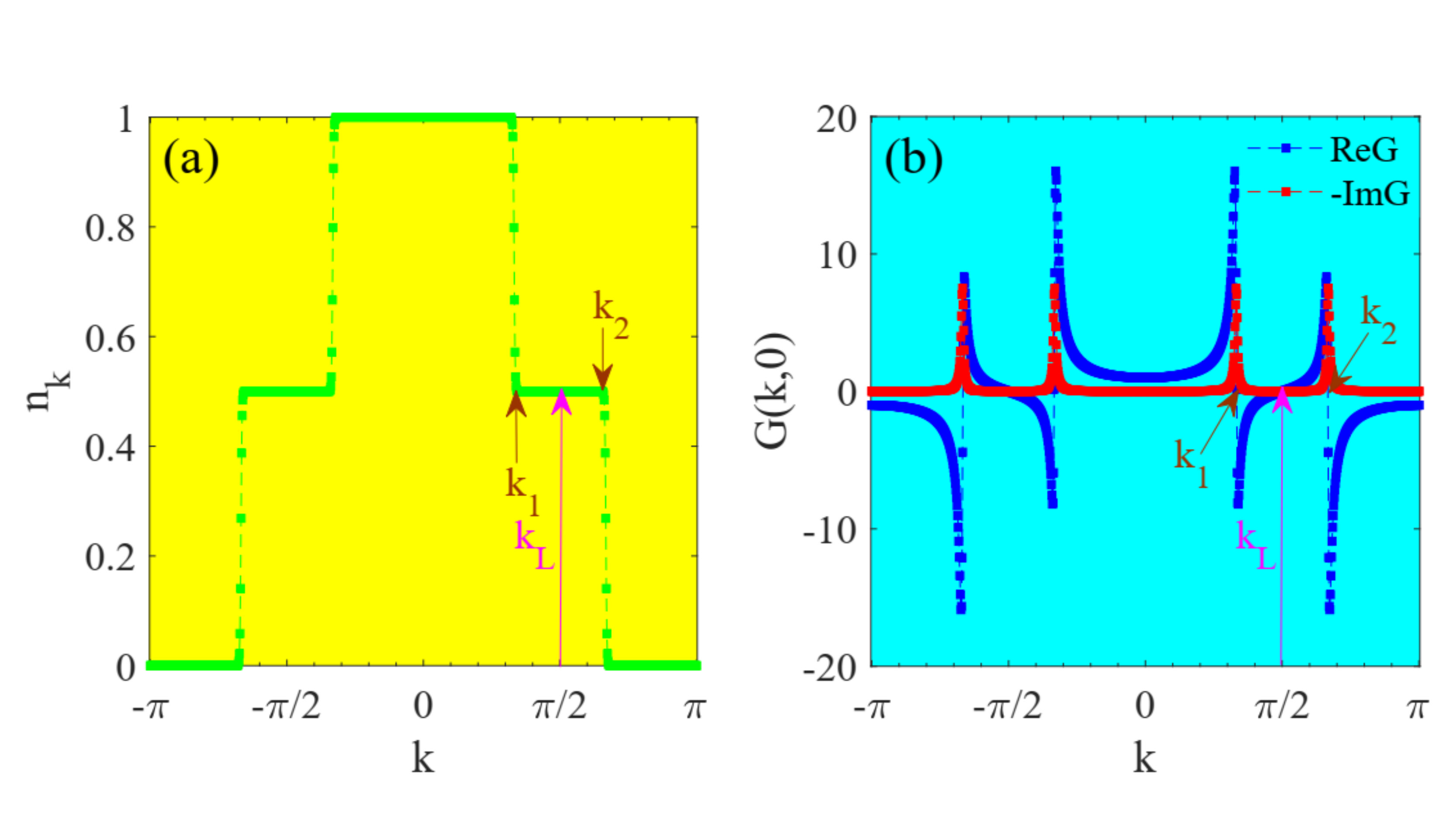}
\caption{\label{fig:kL} (a) Electron's distribution function $n_{k}$ and (b) the real (imaginary) part of single-particle Green's function at zero-frequency ($\mathrm{Re}G(k,0),\mathrm{Im}G(k,0)$) for $U/t=3,\mu=U/2$.}
\end{figure}

However, we should emphasize that although metallic NFL phase in HK shows FO and seems to fit with non-interacting formula $\delta n_{i}\sim \frac{\cos(2k_{F}|R_{i}|)}{|R_{i}|}$, there does not exist well-defined Fermi wave-vector in $k_{F}$. This fact can be seen in Fig.~\ref{fig:kL}(a), where the particle distribution $n_{k}$ (calculated with Eq.~\ref{eq6} at $T=0$) shows FL-like jump at $k_{1}$ and $k_{2}$ but not at putative Fermi wavevector $k_{F}=\pi/2$. The jump at $k_{1},k_{2}$ suggests two-Fermi-surface structure and their location is determined by
\begin{equation}
k_{2}=\left|\arccos\frac{U-\mu}{2t}\right|,~~~~k_{1}=\left|\arccos\frac{-\mu}{2t}\right|,\nonumber
\end{equation}
which results from inspecting Eq.\ref{eq6} at $T=0$ ($n_{k}^{T=0}=\frac{1}{2}\left[\theta(2t\cos k+U/2)+\theta(2t\cos k-U/2)\right]$).
If we focus on regime with $k>0$, the jump at $k_{1}$ ($k_{2}$) corresponds to the occupation of electron from $n_{k\sigma}=1$ to $1/2$ (from $1/2$ to $0$). Furthermore, the real part of Green's function at zero-frequency ($\mathrm{Re}G(k,0)$) diverges at $k_{1},k_{2}$, (Fig.~\ref{fig:kL}(b)) thus, we may call $k_{1},k_{2}$ as quasi-Fermi wavevector.

At the same time, one is able to calculate density of electron in terms of $k_{2},k_{1}$,
\begin{equation}
n=\overbrace{2k_{1}}^{\Omega_{2}}\frac{2}{2\pi}+\overbrace{2(k_{2}-k_{1})}^{\Omega_{1}}\frac{2}{2\pi}\frac{1}{2}=(k_{1}+k_{2})\frac{2}{2\pi}\equiv\frac{1}{\pi}2k_{a}.\nonumber
\end{equation}
where the prefactor $\frac{2}{2\pi}$ denotes the density of state in momentum space with spin degeneracy while the factor $\frac{1}{2}$ uncovers the fact that only single occupation exists in $\Omega_{1}$ and we have $n_{k}=1/2$ in this regime. We have also defined the average Fermi wavevector $k_{a}=(k_{1}+k_{2})/2$, whose effect is just like the Fermi wavevector $k_{F}$ in $1D$ Fermi gas.

Note that, one observes zero point in the real part of Green's function ($\mathrm{Re}G(k,0)$), (Fig.~\ref{fig:kL}(b)) such zero point defines the Luttinger surface instead of more familiar Fermi surface.\cite{Dzyaloshinskii2003} So, the corresponding characteristic wavevector is named Luttinger wavevector $k_{L}$. In our case, we find $k_{L}=\pi/2$ from $\mathrm{Re}G(k=k_{L},0)=0$ ($\mu=U/2$), which is identical to non-interacting Fermi wavevector $k_{F}$ and the average Fermi wavevector $k_{a}$.

Considering $k_{L},k_{a}$, we should identify which one determines the FO in NFL phase of HK,
\begin{equation}
\delta n_{i}^{HK}\sim \frac{\cos(2k_{a}|R_{i}|)}{|R_{i}|},~~~~~~~~|R_{i}|>>1,\nonumber
\end{equation}
or
\begin{equation}
\delta n_{i}^{HK}\sim \frac{\cos(2k_{L}|R_{i}|)}{|R_{i}|},~~~~~~~~|R_{i}|>>1.\nonumber
\end{equation}
\subsection{Linear response theory}
In last subsection, we have seen that in terms of $T$-matrix approximation, metallic NFL state in HK model indeed exhibits FO, but unexpectedly, it seems to be determined by Luttinger wavevector $k_{L}$ or average Fermi wavevector $k_{a}$. To pin down which one is responsible for FO, here, we use linear response theory to estimate the electron density and it may be considered as a crosscheck on previous computation.

According to linear response theory,\cite{Coleman2015} we have
\begin{eqnarray}
\delta n_{i}(t)&=&\frac{1}{i}\int_{-\infty}^{t}dt'\langle[\hat{n}_{i}(t),\hat{H}_{imp}(t')]\rangle\nonumber\\
&=&
\frac{1}{i}\int_{-\infty}^{\infty}dt'\theta(t-t')\langle[\hat{n}_{i}(t),\hat{n}_{0}(t')]\rangle V(t')\nonumber\\
&=&-\int_{-\infty}^{\infty}dt' \chi_{c}(R_{i},R_{0},t-t')V(t')\nonumber
\end{eqnarray}
Here, the charge susceptibility is defined as
\begin{equation}
\chi_{c}(R_{i},R_{0},t-t')=-\frac{1}{i}\theta(t-t')\langle[\hat{n}_{i}(t),\hat{n}_{0}(t')]\rangle.\nonumber
\end{equation}
Thus, if we are able to calculate $\chi_{c}(R_{i},R_{0},t-t')$, the electron density is easy to obtain after integrating over time.
Equivalently, $\delta n_{i}(t)=-\int \frac{d\omega}{2\pi}e^{-i\omega t}\chi_{c}(R_{i},R_{0},\omega)V(\omega)$
and $\chi_{c}(R_{i},R_{0},\omega)=\int dt e^{i\omega t}\chi_{c}(R_{i},R_{0},t)$.

In many-body physics, one always uses Wick rotation and calculates imaginary-time charge susceptibility
\begin{equation}
\chi_{c}(R_{i},R_{0},\tau)=\langle\hat{T}_{\tau}\hat{n}_{i}(\tau)\hat{n}_{0}\rangle,\nonumber
\end{equation}
or its Fourier transformation $\chi_{c}(R_{i},R_{0},i\Omega_{n})$. Finally, we obtain $\chi_{c}(R_{i},R_{0},\omega)=\chi_{c}(R_{i},R_{0},i\Omega_{n}\rightarrow \omega+i0^{+})$.

In the framework of perturbation theory with the help of Feynman diagrams, using Wick theorem, one can calculate $\chi_{c}(R_{i},R_{0},i\Omega_{n})$ or its Fourier transformation $\chi_{c}(q,i\Omega_{n})$ easily. When interaction is turned off ($U=0$), we just obtain (see Appendix.~\ref{ap_B})
\begin{equation}
\chi_{c}^{(0)}(q,i\Omega_{n})=\frac{-1}{N_{s}\beta}\sum_{k,\omega_{n}}\sum_{\sigma}G_{\sigma}^{(0)}(k+q,\omega_{n}+\Omega_{n})G_{\sigma}^{(0)}(k,\omega_{n}).\label{eq:10}
\end{equation}
After frequency summation, one finds the standard result $\chi_{c}^{(0)}(q,i\Omega_{n})=\frac{2}{N_{s}}\sum_{k}\frac{f_{F}(\varepsilon_{k+q}-\mu)-f_{F}(\varepsilon_{k}-\mu)}{i\Omega_{n}-\varepsilon_{k+q}+\varepsilon_{k}}.$

However, for HK model, we have not noticed any perturbation theory which can reproduce the exact solution like Green's function or free energy. (Note however that Ref.~\onlinecite{Wang2023} has proposed a Hartree-Fock based perturbation theory for HK model at $T=0$.) Therefore, one should be careful when calculating multi-particle correlation like $\chi_{c}$.

Fortunately, Ref.~\onlinecite{Phillips2018} and \onlinecite{Nogueira1996} tell us that, the charge and spin susceptibility of HK has identical formalism to familiar non-interacting electron gas (e.g. Eq.~\ref{eq:10}) and the only difference is to replace the non-interacting Green's function with the interacting one (Eq.~\ref{eq4}). For HK model, we just replace $G_{\sigma}^{(0)}$ with $G_{\sigma}$ and the explicit result is given by Ref.~\onlinecite{Phillips2018} as
\begin{eqnarray}
\chi_{c}(q,i\Omega_{n})&=&\frac{-1}{N_{s}}\sum_{k,\sigma}(1-n_{k})(1-n_{k+q})\nonumber\\
&\times&\frac{f_{F}(\varepsilon_{k}-\mu)-f_{F}(\varepsilon_{k+q}-\mu)}{i\Omega_{n}-\varepsilon_{k+q}+\varepsilon_{k}}\nonumber\\
&+&\frac{-1}{N_{s}}\sum_{k,\sigma}(1-n_{k})n_{k+q}\nonumber\\
&\times&\frac{f_{F}(\varepsilon_{k}-\mu)-f_{F}(\varepsilon_{k+q}-\mu+U)}{i\Omega_{n}-\varepsilon_{k+q}-U+\varepsilon_{k}}\nonumber\\
&+&\frac{-1}{N_{s}}\sum_{k,\sigma}n_{k}(1-n_{k+q})\nonumber\\
&\times&\frac{f_{F}(\varepsilon_{k}-\mu+U)-f_{F}(\varepsilon_{k+q}-\mu)}{i\Omega_{n}-\varepsilon_{k+q}+\varepsilon_{k}+U}\nonumber\\
&+&\frac{-1}{N_{s}}\sum_{k,\sigma}n_{k}n_{k+q}\nonumber\\
&\times&\frac{f_{F}(\varepsilon_{k}-\mu+U)-f_{F}(\varepsilon_{k+q}-\mu+U)}{i\Omega_{n}-\varepsilon_{k+q}+\varepsilon_{k}}\label{eq:28}
\end{eqnarray}
It is noted that the above density-density correlation function is not an approximation but is exact for HK model due to the solvability.
It is easy to check that when $U=0$, the above result reduces into the non-interacting one $\chi_{c}^{(0)}$.

\begin{figure}
\includegraphics[width=0.9\linewidth]{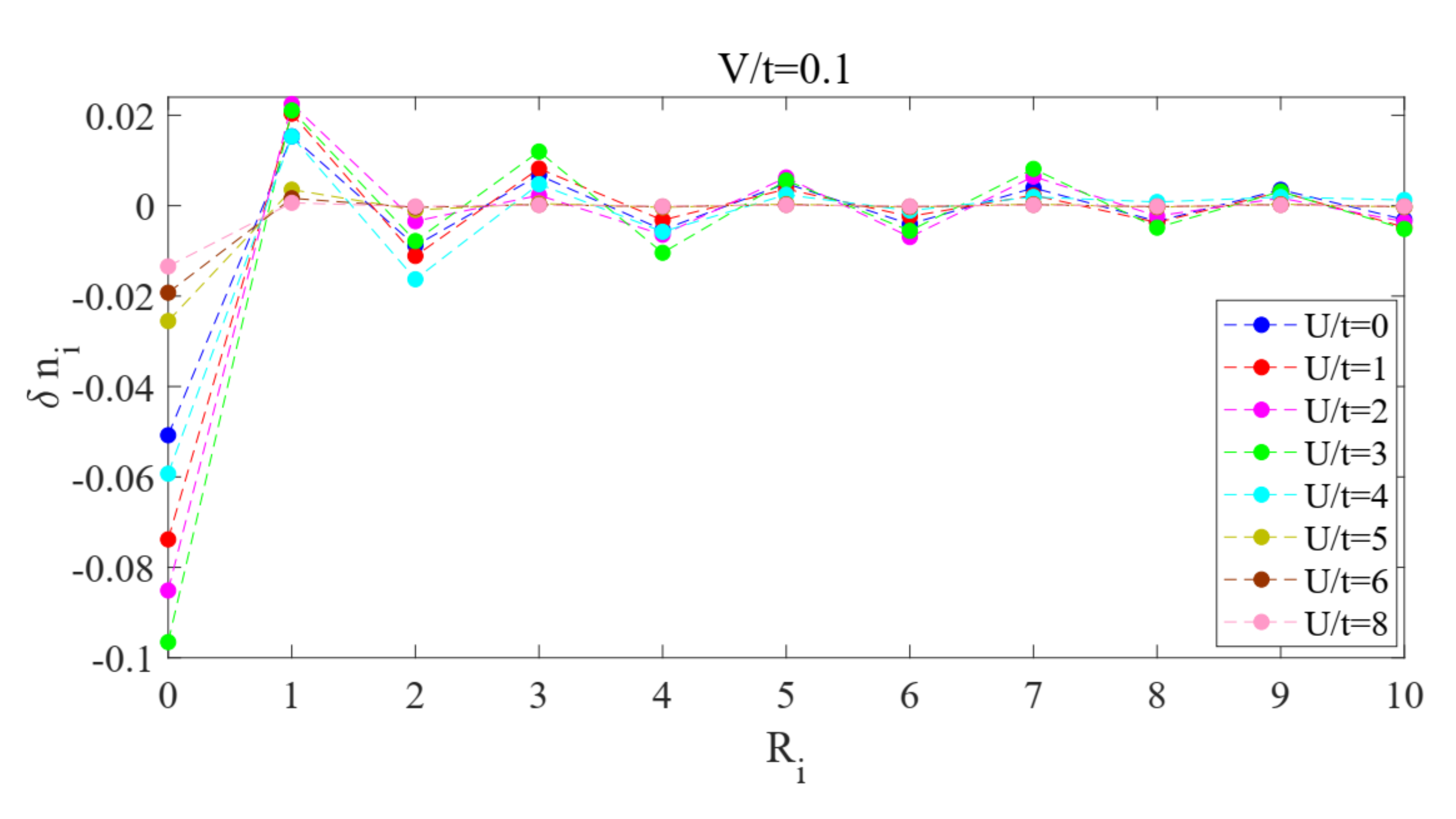}
\caption{\label{fig:V_01_LRT} $\delta n_{i}$ calculated from the linear response theory for $U/t=0,1,2,3,4,5,6,8$ with $V/t=0.1$ and $\mu=U/2$.}
\end{figure}

Since the strength of impurity is static, we should have $V(\omega)=2\pi V\delta(\omega)$, so $\delta n_{i}=-V\chi_{c}(R_{i},R_{0},\omega=0)$. This can be calculated if we replace $i\Omega_{n}$ with $\omega+i0^{+}$ in Eq.~\ref{eq:28} and make a Fourier transformation,
\begin{equation}
\delta n_{i}=-V\frac{1}{N_{s}}\mathrm{Re}\left[\sum_{q}e^{iq(R_{i}-R_{0})}\chi_{c}(q,i\Omega_{n}\rightarrow\omega+i0^{+})|_{\omega=0}\right].
\end{equation}
In Fig.~\ref{fig:V_01_LRT}, we have plotted the results from the linear response theory and good agreement with our previous $T$-matrix calculation (Fig.~\ref{fig:V_01}) has been found.

\begin{figure}
\includegraphics[width=0.9\linewidth]{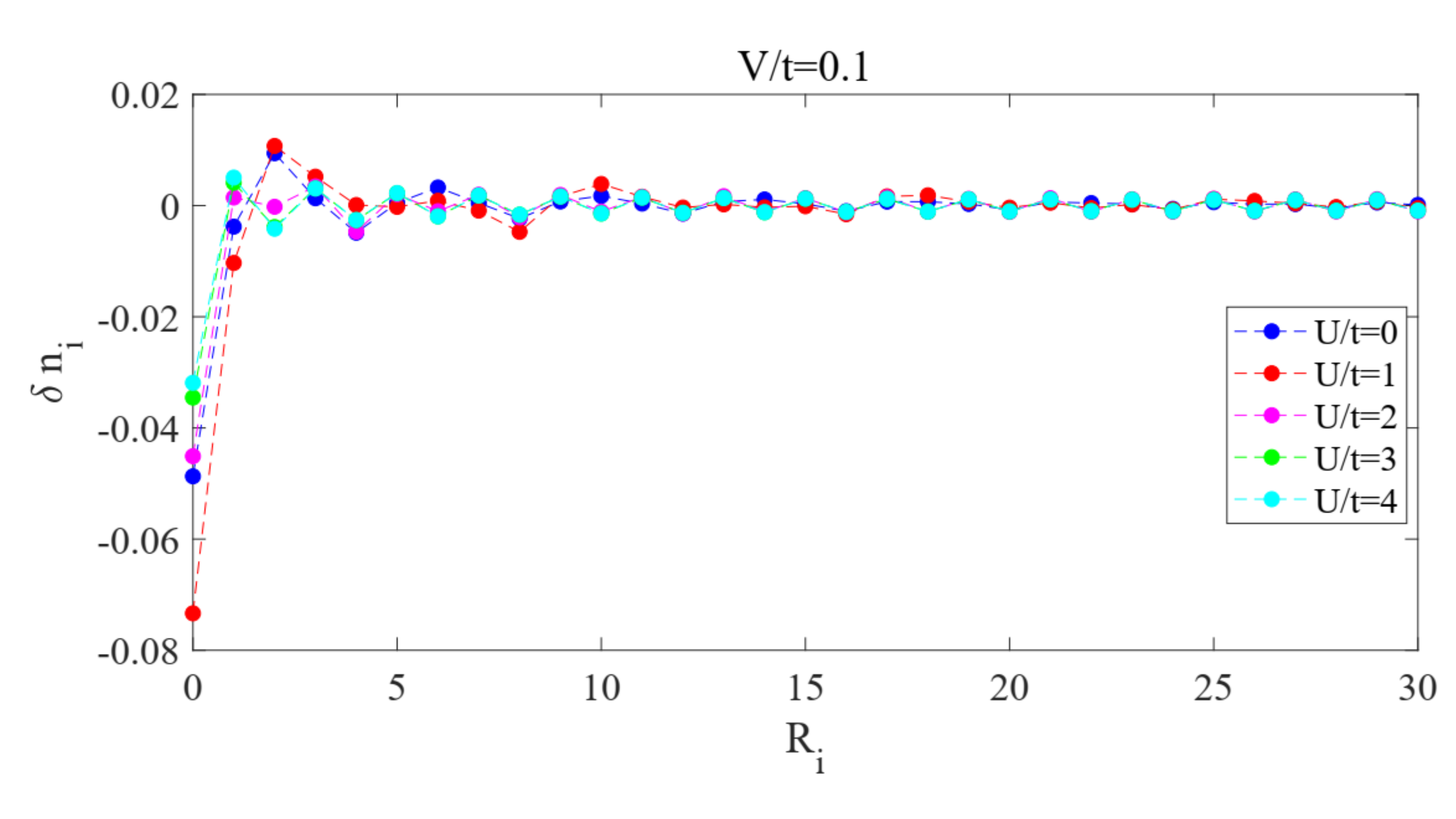}
\caption{\label{fig:V_01_LRT_05} $\delta n_{i}$ for $U/t=0,1,2,3,4$ with $V/t=0.1$ and electron density is fixed to $0.5$.}
\end{figure}

\begin{figure}
\includegraphics[width=1.1\linewidth]{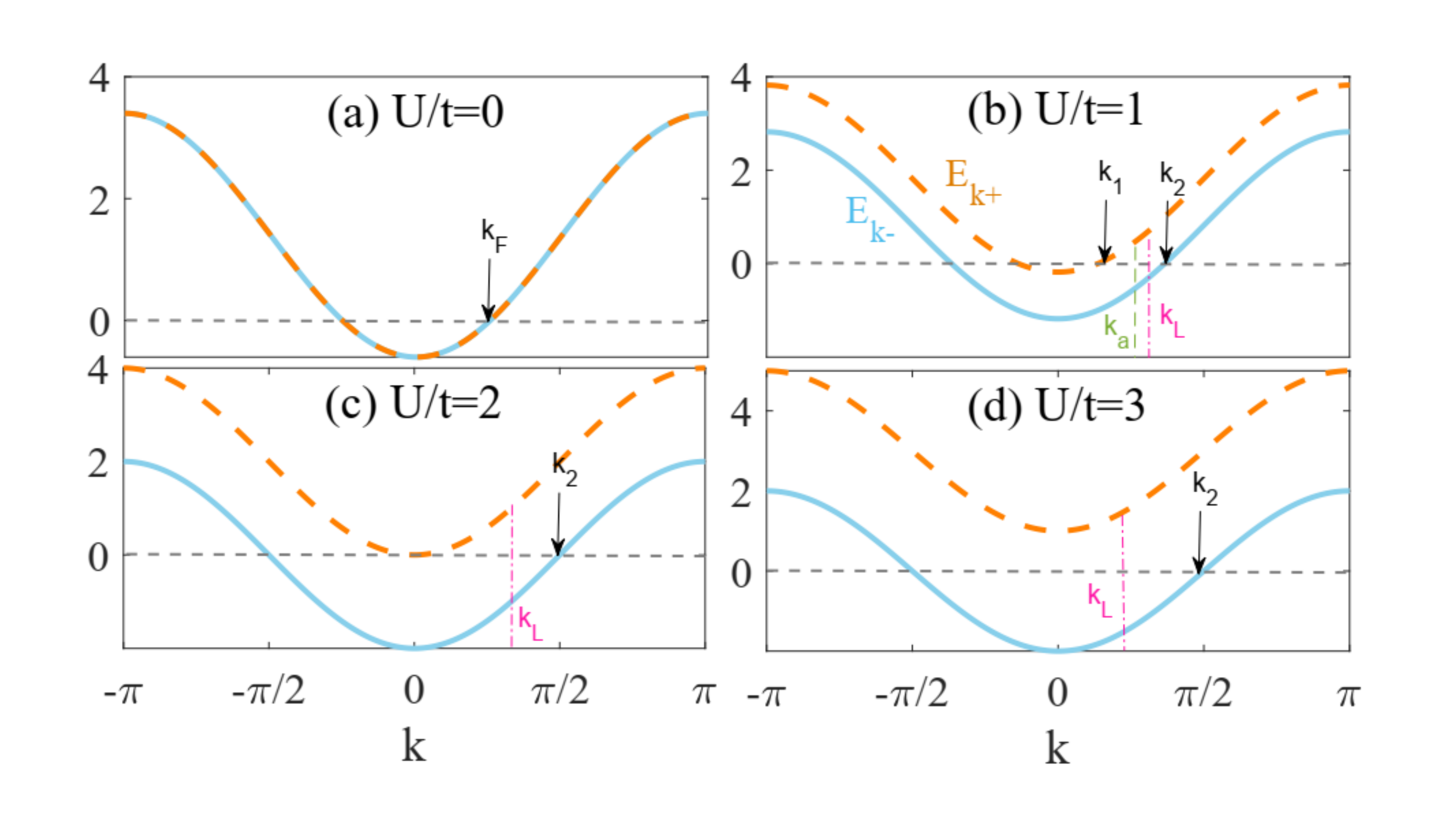}
\caption{\label{fig:band} The quasi-particle bands $E_{k-}=\varepsilon_{k}-\mu,E_{k+}=\varepsilon_{k}-\mu+U$ for $U/t=0,1,2,3$ with electron density $n=0.5$.}
\end{figure}

\begin{figure}
\includegraphics[width=1.0\linewidth]{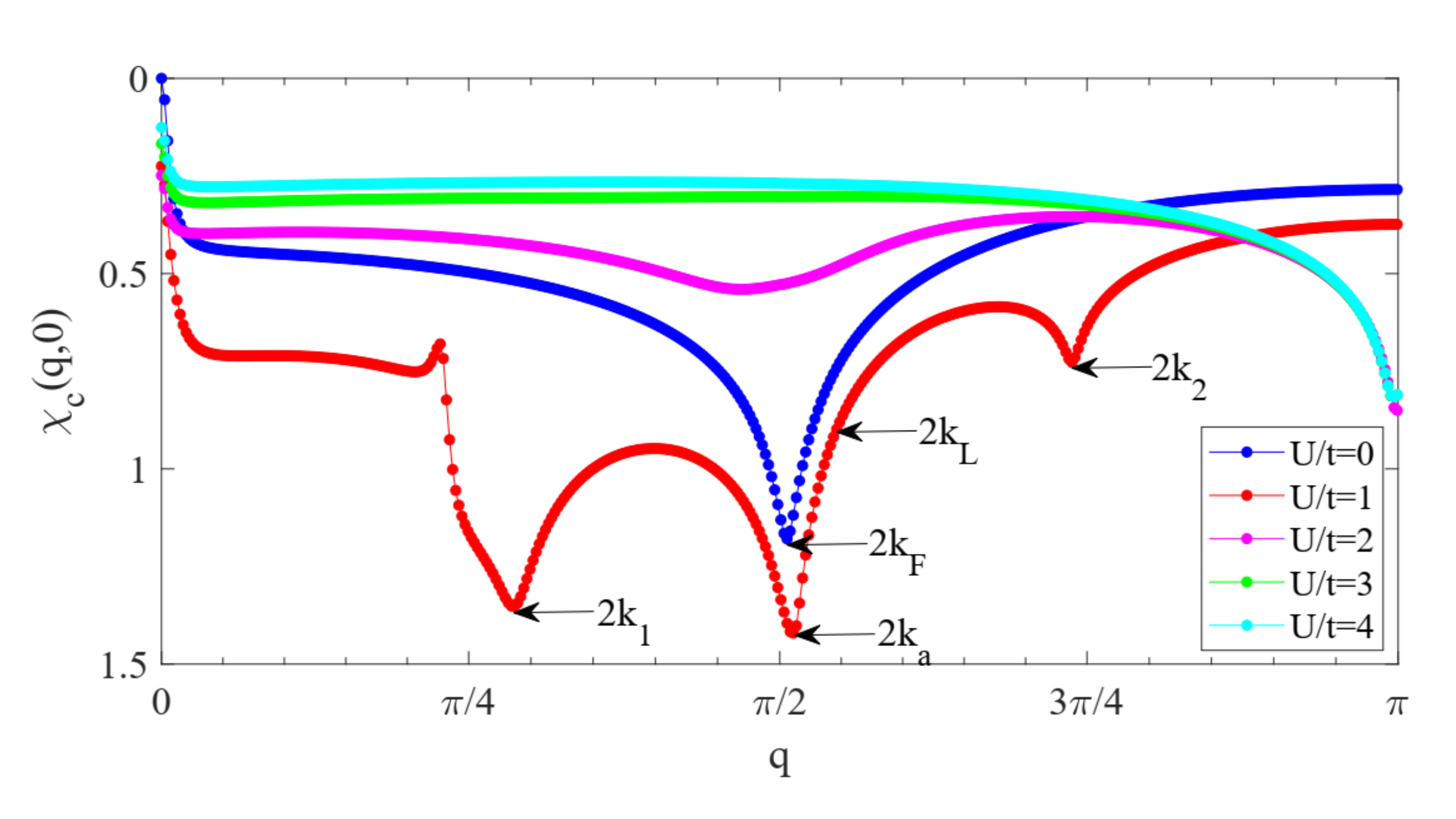}
\caption{\label{fig:V_01_chi} $\chi_{c}(q,0)$ for $U/t=0,1,2,3,4$ with $V/t=0.1$ and electron density is fixed to $0.5$.}
\end{figure}

Because, $k_{L}$ and $k_{a}$ are identical in the symmetric case $\mu=U/2$. Instead, we investigate the asymmetric case, e.g. the situation with electron density $n=0.5$ in Fig.~\ref{fig:V_01_LRT_05}. (Results for electron density $n=0.4$ and $n=0.6$ are shown Appendix.~\ref{ap_C} and no physics is changed.) For this electron density, certain regime of quasi-particle bands $E_{k-}=\varepsilon_{k}-\mu,E_{k+}=\varepsilon_{k}-\mu+U$ are doubly occupied when $U/t$ is small, in contrast, only $E_{k-}$ band will be occupied when $U/t$ is large. (See corresponding band structure $E_{k\pm}$ in Fig.~\ref{fig:band}.) Here, although FO is still visible but it is not easy to distinguish $k_{L}$ and $k_{a}$, thus we plot $\chi_{c}(q,0)$ ($\chi_{c}(q,0)\propto \delta n(q)$, the Fourier transformation of $\delta n_{i}$) in Fig.~\ref{fig:V_01_chi}, which is able to show the dominating wavevector for charge response.

As one can see from Fig.~\ref{fig:V_01_chi}, when $U/t=0$, the correct $2k_{F}$ singularity of non-interacting electron gas is reproduced as expected. This means we can approximate $\chi_{c}(q)$ as $\chi_{c}(2k_{F})$ and it leads to $\delta n_{i}\sim\cos(2k_{F}(R_{i}-R_{0}))\chi_{c}(2k_{F})$. Thus, the correct $2k_{F}$ oscillation of $\delta n_{i}$ has been reproduced.

Next, if one enhances the interaction with $U/t=1$, three dominating peaks located in $2k_{1},2k_{2}$ and $2k_{a}$ are found. In this case,
\begin{eqnarray}
\delta n_{i}&\sim&\cos(2k_{1}(R_{i}-R_{0}))\chi_{c}(2k_{1})
+\cos(2k_{2}(R_{i}-R_{0}))\chi_{c}(2k_{2})\nonumber\\
&+&\cos(2k_{a}(R_{i}-R_{0}))\chi_{c}(2k_{a}), \nonumber
\end{eqnarray}
and the last one with $2k_{a}$ singularity appears to be the final winner. (We will provide an intuitive explanation later.)

In addition, if $U$ is larger, $\chi_{c}(q,0)$ has peak near $\pi$, which is just $2k_{2}$ since in this case, only one quasi-particle band $E_{k-}$ has been occupied.

Actually, as can be seen in Fig.~\ref{fig:band}(b), one may consider $2k_{a}=k_{1}+k_{2}$ as the inter-band particle-hole excitation (scattering) process induced by the HK interaction. In other words, this can be explained as the inter-band '$2k_{F}$' singularity, which is in contrast with the intra-band '$2k_{F}$' singularity like $2k_{2}$ in Fig.~\ref{fig:band}(d). At the same time, we see that in Eq.~\ref{eq:28}, the first and the fourth term are mainly dominated by intra-band particle-hole excitation while the second and the third term show inter-band particle-hole excitation. If one considers the numerator in Eq.~\ref{eq:28}, e.g. the prefactor of first term $(1-n_{k})(1-n_{k+q})(f_{F}(\varepsilon_{k}-\mu)-f_{F}(\varepsilon_{k+q}-\mu))$ and the prefactor of third term $n_{k}(1-n_{k+q})(f_{F}(\varepsilon_{k}-\mu+U)-f_{F}(\varepsilon_{k+q}-\mu))$, it is found that the former one is more mismatched than the latter one, thus the inter-band excitation (described by the second and third term in Eq.~\ref{eq:28}) wins over the intra-band excitation (the first and fourth term in Eq.~\ref{eq:28}). So, we expect that the charge susceptibility is determined by the average Fermi surface, which highlights the inter-band particle-hole excitation.

Therefore, we should conclude that FO in NFL state with two Fermi surface is determined by the average Fermi surface while NFL state with only one Fermi surface behaves like usual $2k_{F}$ singularity.

\subsection{FO at finite temperature}
\begin{figure}
\includegraphics[width=1.15\linewidth]{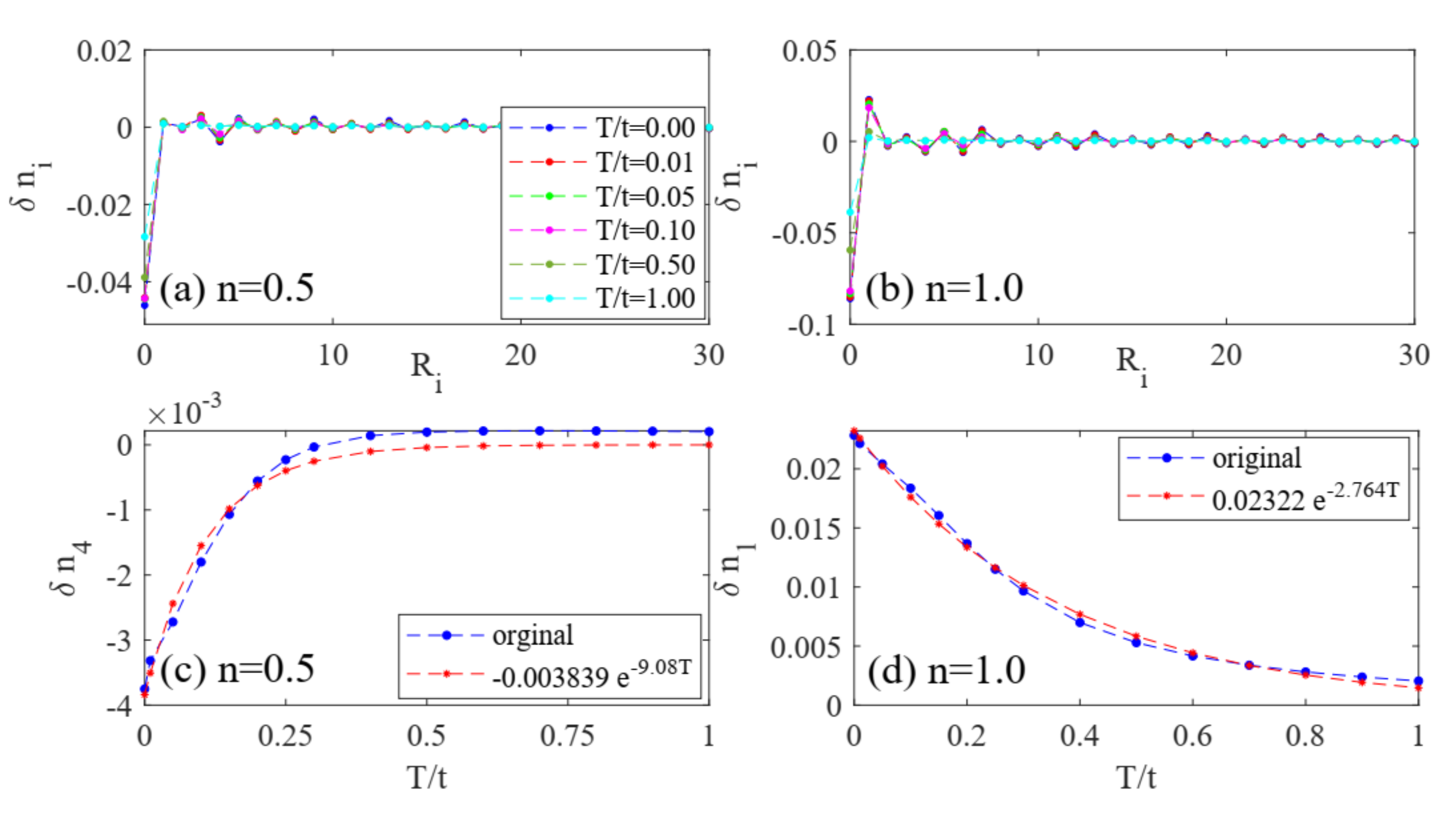}
\caption{\label{fig:T} $\delta n_{i}$ versus temperature for (a) $n=0.5$ and (b) $n=1.0$ with $V/t=0.1,U/t=2$.
(c) $\delta n_{4}$ denotes amplitude of FO on the fourth nearest-neighbor site and (d) $\delta n_{1}$ is for the nearest-neighbor site around impurity.}
\end{figure}
Before ending this section, we try to explore the finite temperature effect of FO. It is expected that the thermal effect will wash out the sharp jump around Fermi surface, such that the charge response will be weakened if elevating temperature. Thus, the amplitude of FO decreases when temperature is increased, which can be seen in Fig.~\ref{fig:T}. Here, $\delta n_{i}$ for $n=0.5$ and $n=1$ are shown in Fig.~\ref{fig:T}(a) and (b).
We see that when $T/t\lesssim 0.1$, the amplitude of FO is visible while higher temperature does not lead to noticeable FO. The amplitude of FO has also been fitted with exponential function $\sim e^{-T}$ and good agreement with the calculated ones has been found Fig.~\ref{fig:T}(c) and (d).

\section{Discussion}\label{sec3}
\subsection{FO on $2D$ square lattice}
\begin{figure}
\includegraphics[width=1.00\linewidth]{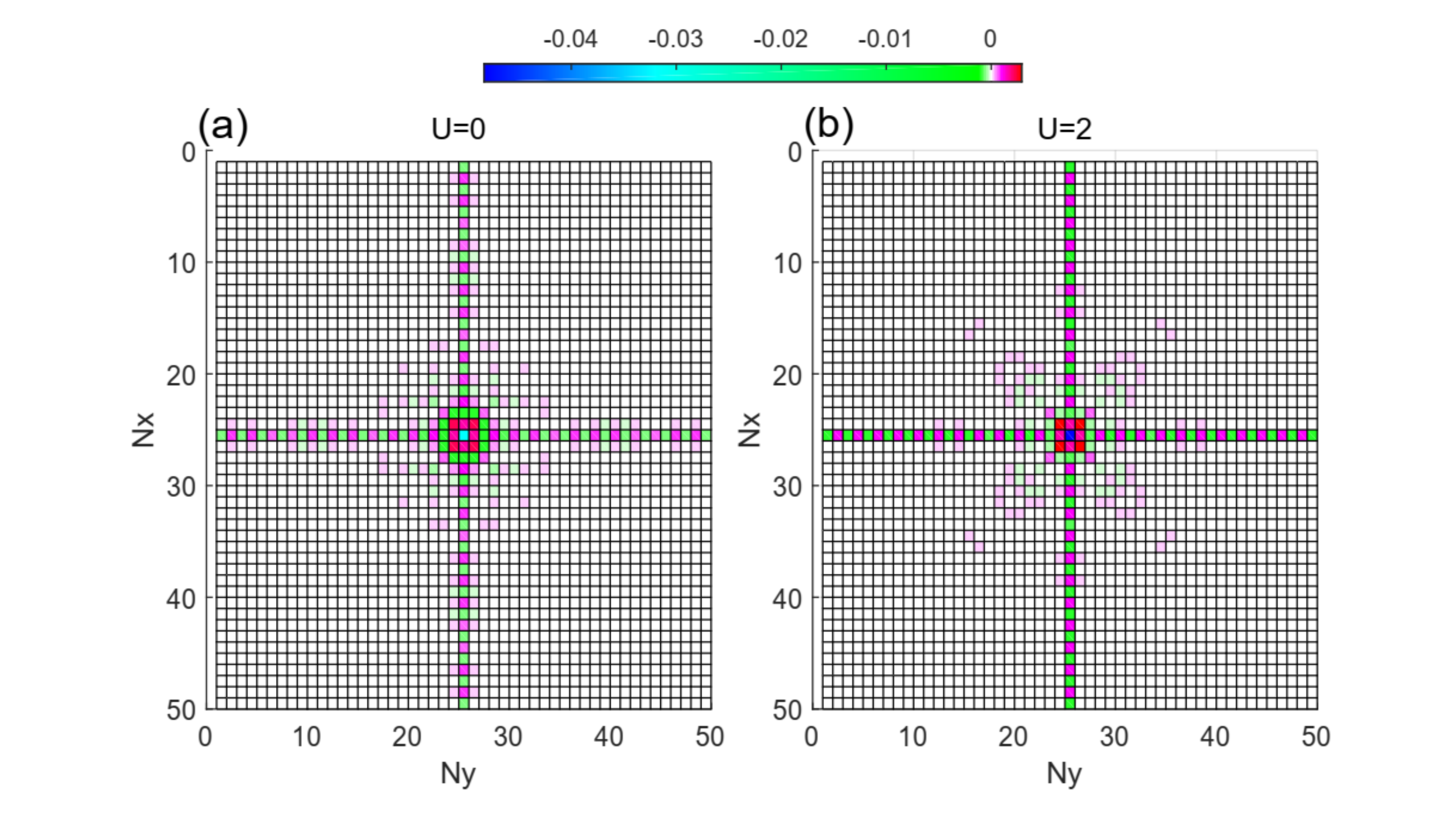}
\caption{\label{fig:2d_1} $\delta n_{i}$ on square lattice for (a) $U/t=0$, (b) $U/t=2$ with $V/t=0.1$ and electron density is fixed to $0.5$.}
\end{figure}

\begin{figure}
\includegraphics[width=1.40\linewidth]{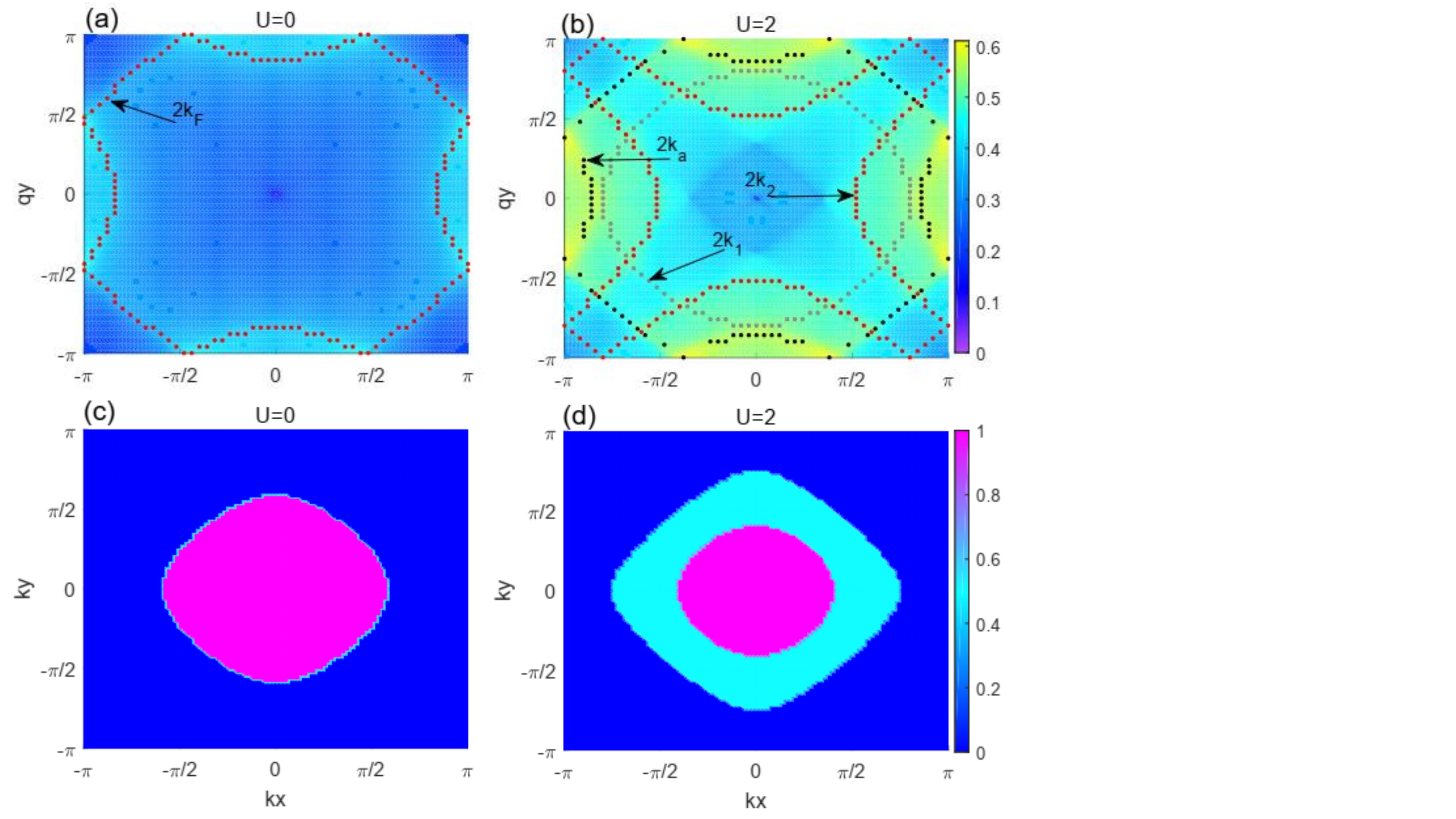}
\caption{\label{fig:2d_2} $\chi_{c}(q,0)$ (a)(b) and $n_{k}$ (c)(d) on square lattice for $U/t=0$ and $U/t=2$ with $V/t=0.1$ and electron density is fixed to $0.5$.}
\end{figure}

Now, we turn our attention to the $2D$ square lattice, where the dispersion of electron is chosen to be $\varepsilon_{k}=-2t(\cos k_{x}+\cos k_{y})$. Generally, the findings in the $1D$ situation are still valid in the present $2D$ square lattice as can be seen in Figs.~\ref{fig:2d_1} and \ref{fig:2d_2}, which is not unexpected since the NFL states in HK model belong to the same universality class and the effect of space dimension does not change the nature of NFL. This feature is quite different from the case in Hubbard model, where NFL in $1D$ is faithfully described by the Luttinger liquid paradigm, however its extension to the most important $2D$ case is still lacking. In our opinion, such a difference may result from  competing symmetry-breaking states involving charge-density, spin-density and pairing order in Hubbard model and including these orders for HK-like model is able to clarify this issue.

Fig.~\ref{fig:2d_1} shows $\delta n_{i}$ for $U/t=0$ and $U/t=2$, and FO exists in these two situations despite of the latter one being a NFL. Furthermore, in Fig.~\ref{fig:2d_2}(b), it is seen that $\chi_{c}(q,0)$ in NFL is dominated by the average Fermi surface ($2k_{a}$ in the figure) as expected. As comparison the non-interacting Fermi gas in Fig.~\ref{fig:2d_2}(a) has the usual $2k_{F}$ charge response. In addition, the
two-Fermi-surface structure of quasi-particle in NFL exists in Fig.~\ref{fig:2d_2}(d).
\subsection{More impurities?}
In previous sections, we have studied the details of the single impurity problem and the clear signature of FO is shown in metallic NFL states. But, if more impurities are involved, what will happen? It is expected that if the density of impurities is small, the interference effect between impurities can be safely ignored and one just uses the picture of the single impurity problem. However, if more impurities exist, interference effect must be included, which invalidates the T-matrix formalism we have developed for single impurity case. Although we cannot make any definite prediction due to the lack of appropriate theoretical formalism, it seems that localization of electron (Anderson localization) is an inevitable one for strong impurity scattering. Future study on the interplay between localization and electron correlation in HK-like models is desirable and it may relate to the timely issue of quantum thermalization or many-body localization.\cite{Abanin2019}

\subsection{Magnetic impurity}
In the main text, only effect from the nonmagnetic impurity is analyzed. However, we all know that the understanding on magnetic impurity is an essential issue in condensed matter physics, such as Kondo effect, Ruderman-Kittel-Kasuya-Yosida exchange interaction and the spin glass state.\cite{Hewson1993,Contucci2013} It is noted that the former one has already been explored by poorman's scaling approach and deviation from usual Kondo impurity in FL is noticeable.\cite{Setty2021}

Due to the perturbative nature of poorman's scaling, the ground-state of Kondo impurity in HK model has not been established and it seems that the state-of-art numerical renormalization group, which is very successful on Kondo impurity in non-interacting environment,\cite{Bulla2008} cannot be utilized without nontrivial modification. To explore the ground-state, it will be helpful to follow the classic variational wave-function calculation of Yosida as the first step.\cite{Yosida1966}
\section{Conclusion and Future direction}\label{sec4}
In conclusion, we have found that Friedel oscillation exists in non-Fermi liquid phase in the HK model, which results from the calculation of $T$-matrix approximation and linear response theory. When there exits two-Fermi-surface structure, inter-band particle-hole excitation dominates and one observes the average Fermi surface. We should emphasize that the two-Fermi-surface structure in HK model is an intrinsic effect induced by interaction and no symmetry breaking is involved. This is in contrast with the usual multi-band system, in which the bands can appear without electron correlation.

In fact, besides the HK model, the average Fermi surface structure may naturally arise in many correlated electron systems, e.g. the phenomenological description of underdoped cuprate in terms of Yang-Rice-Zhang ansatz,\cite{Rice2012} Hubbard-I approximation solution of Hubbard model, Falicov-Kimball model and Ising-Kondo lattice.\cite{Hubbard1963,wwYang2022,wwYang2019,wwYang2021} Thus, it is interesting to examine whether the finding in this work is still valid in those more realistic systems, which contributes to our understanding on high-temperature cuprate superconductivity.

For future study, considering the recent progress on superconductivity in HK-like models,\cite{Phillips2020,Zhu2021,Zhao2022,Li2022} it will be interesting to explore the impurity effect in those superconducting phases in terms of the framework developed in this work. Since the superconducting phases in HK models are rather different from the usual Bardeen-Cooper-Schrieffer pairing state, we expect that impurity effect will be a good guide to identify the mentioned ones.

Therefore, we hope our work here provides a useful framework to understand Friedel oscillation and related impurity effect in exotic correlated electron models like HK model. Certain extensions of our work will contribute to the exploration of impurity effect in generic strongly correlated electron systems.

\section*{Acknowledgments}
We acknowledge funding from the National Key Research and Development Program of China (Grant No.2022YFA1402704) and the National Natural Science Foundation of China (Grant No.12047501 and No.11834005).

\appendix
\section{Derivation of singe-particle Green's function}\label{ap_A}
Follow the treatment of Hubbard model,\cite{Hubbard1963} let us define the single-particle Green's function as $G_{\sigma}(k,\omega)=\langle\langle \hat{c}_{k\sigma}|\hat{c}_{k\sigma}^{\dag}\rangle\rangle_{\omega}$, which is just the Fourier transformation of the retarded Green's function
\begin{equation}
G_{\sigma}(k,t)=-i\theta(t)\langle[\hat{c}_{k\sigma}(t),\hat{c}_{k\sigma}^{\dag}]_{+}\rangle.\nonumber
\end{equation}
Then, in terms of
\begin{eqnarray}
&&[\hat{c}_{k\sigma},\hat{H}]=(\varepsilon_{k}-\mu)\hat{c}_{k\sigma}+U\hat{c}_{k\sigma}\hat{n}_{k\bar{\sigma}},\nonumber\\
&&[\hat{c}_{k\sigma}\hat{n}_{k\bar{\sigma}},\hat{H}]=(\varepsilon_{k}-\mu+U)\hat{c}_{k\sigma}\hat{n}_{k\bar{\sigma}},\nonumber
\end{eqnarray}
we find
\begin{equation}
\omega\langle\langle \hat{c}_{k\sigma}|\hat{c}_{k\sigma}^{\dag}\rangle\rangle_{\omega}=1+(\varepsilon_{k}-\mu)\langle\langle \hat{c}_{k\sigma}|\hat{c}_{k\sigma}^{\dag}\rangle\rangle_{\omega}+U\langle\langle \hat{c}_{k\sigma}\hat{n}_{k\bar{\sigma}}|\hat{c}_{k\sigma}^{\dag}\rangle\rangle_{\omega}\nonumber
\end{equation}
and
\begin{equation}
\omega\langle\langle \hat{c}_{k\sigma}\hat{n}_{k\bar{\sigma}}|\hat{c}_{k\sigma}^{\dag}\rangle\rangle_{\omega}=\langle \hat{n}_{k\bar{\sigma}}\rangle+(\varepsilon_{k}-\mu+U)\langle\langle \hat{c}_{k\sigma}\hat{n}_{k\bar{\sigma}}|\hat{c}_{k\sigma}^{\dag}\rangle\rangle_{\omega}\nonumber
\end{equation}

Since above equations are closed, we obtain
\begin{equation}
\langle\langle \hat{c}_{k\sigma}\hat{n}_{k\bar{\sigma}}|\hat{c}_{k\sigma}^{\dag}\rangle\rangle_{\omega}=\frac{\langle \hat{n}_{k\bar{\sigma}}\rangle}{\omega-\varepsilon_{k}+\mu-U}\nonumber
\end{equation}
and
\begin{eqnarray}
G_{\sigma}(k,\omega)&=&\frac{1+\frac{U\langle\hat{n}_{k\bar{\sigma}}\rangle}{\omega-(\varepsilon_{k}-\mu+U)}}{\omega-(\varepsilon_{k}-\mu)}\nonumber\\
&=&\frac{1-\langle\hat{n}_{k\bar{\sigma}}\rangle}{\omega-(\varepsilon_{k}-\mu)}+\frac{\langle\hat{n}_{k\bar{\sigma}}\rangle}{\omega-(\varepsilon_{k}-\mu+U)}\nonumber
\end{eqnarray}
which is just the wanted Eq.~\ref{eq4} in the main text.
\section{Charge susceptibility of non-interacting electron}\label{ap_B}
For self-content, we derive the non-interacting formula for charge susceptibility as follows. Firstly, we transformation  $\chi_{c}(R_{i},R_{0},\tau)$ into momentum-energy space via
\begin{equation}
\chi_{c}^{(0)}(R_{i},R_{0},\tau)=\frac{1}{\beta N_{s}}\sum_{\Omega_{n}}\sum_{q}e^{iq(R_{i}-R_{0})-i\Omega_{n}\tau}\chi_{c}^{(0)}(q,\Omega_{n}).
\end{equation}
Then, it is straightforward to derive
\begin{eqnarray}
\chi_{c}^{(0)}(q,\Omega_{n})&=&\int_{0}^{\beta}d\tau e^{i\Omega_{n}\tau}\sum_{j}e^{-iqR_{j}}\chi_{c}^{(0)}(R_{j},0,\tau)\nonumber\\
&=&\sum_{\sigma,\sigma'}\int_{0}^{\beta}d\tau e^{i\Omega_{n}\tau}\sum_{j}e^{-iqR_{j}}\nonumber\\
&\times&\langle\hat{T}_{\tau}\hat{c}_{j\sigma}^{\dag}(\tau)\hat{c}_{j\sigma}(\tau)\hat{c}_{0\sigma'}^{\dag}\hat{c}_{0\sigma'}\rangle\nonumber\\
&=&\frac{1}{N_{s}^{2}}\sum_{k_{1},k_{2},k_{3},k_{4}}\sum_{\sigma,\sigma'}\int_{0}^{\beta}d\tau e^{i\Omega_{n}\tau}\sum_{j}e^{-iqR_{j}}\nonumber\\
&\times&e^{-ik_{1}R_{j}}e^{ik_{2}R_{j}}\langle\hat{T}_{\tau}\hat{c}_{k_{1}\sigma}^{\dag}(\tau)\hat{c}_{k_{2}\sigma}(\tau)\hat{c}_{k_{3}\sigma'}^{\dag}\hat{c}_{k_{4}\sigma'}\rangle\nonumber
\end{eqnarray}
Then, using the standard Wick theorem, we find
\begin{eqnarray}
\chi_{c}^{(0)}(q,\Omega_{n})&=&\frac{-1}{N_{s}}\sum_{k_{1},\sigma}\int_{0}^{\beta}d\tau e^{i\Omega_{n}\tau}G_{\sigma}^{(0)}(k_{1}+q,\tau)G_{\sigma}^{(0)}(k_{1},-\tau)\nonumber\\
&=&\frac{-1}{N_{s}\beta}\sum_{k_{1},\omega_{n},\sigma}G_{\sigma}^{(0)}(k_{1}+q,\omega_{n}+\Omega_{n})G_{\sigma}^{(0)}(k_{1},\omega_{n})\nonumber\\
\label{eq:27}
\end{eqnarray}
Now, if we use the free electron Green's function $G_{\sigma}^{(0)}(k,\omega_{n})=\frac{1}{i\omega_{n}-(\varepsilon_{k}-\mu)}$, it is found that
\begin{eqnarray}
\chi_{c}^{(0)}(q,\Omega_{n})
&=&\frac{-2}{N_{s}\beta}\sum_{k_{1},\omega_{n}}\frac{1}{i\omega_{n}-(\varepsilon_{k}-\mu)}\nonumber\\
&\times&\frac{1}{i\omega_{n}+i\Omega_{n}-(\varepsilon_{k+q}-\mu)}\nonumber\\
&=&\frac{2}{N_{s}}\sum_{k_{1}}\frac{f_{F}(\varepsilon_{k+q}-\mu)-f_{F}(\varepsilon_{k}-\mu)}{i\Omega_{n}-\varepsilon_{k+q}+\varepsilon_{k}}\nonumber.
\end{eqnarray}
which is the textbook result for non-interacting electron gas.
\begin{figure}
\includegraphics[width=0.9\linewidth]{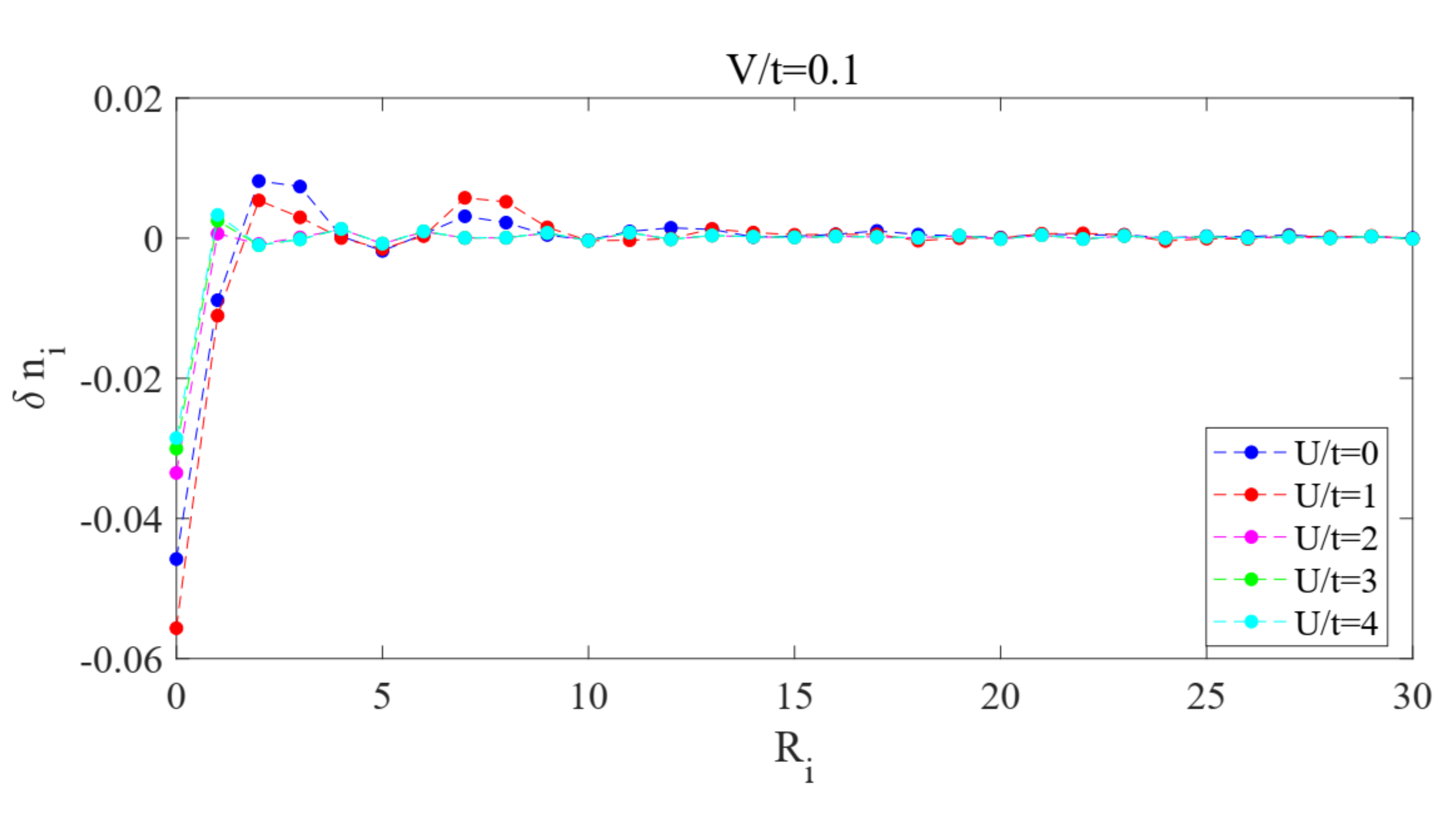}
\caption{\label{fig:V_01_LRT_04} $\delta n_{i}$ for $U/t=0,1,2,3,4$ with $V/t=0.1$ and electron density is fixed to $0.4$.}
\end{figure}
\begin{figure}
\includegraphics[width=0.9\linewidth]{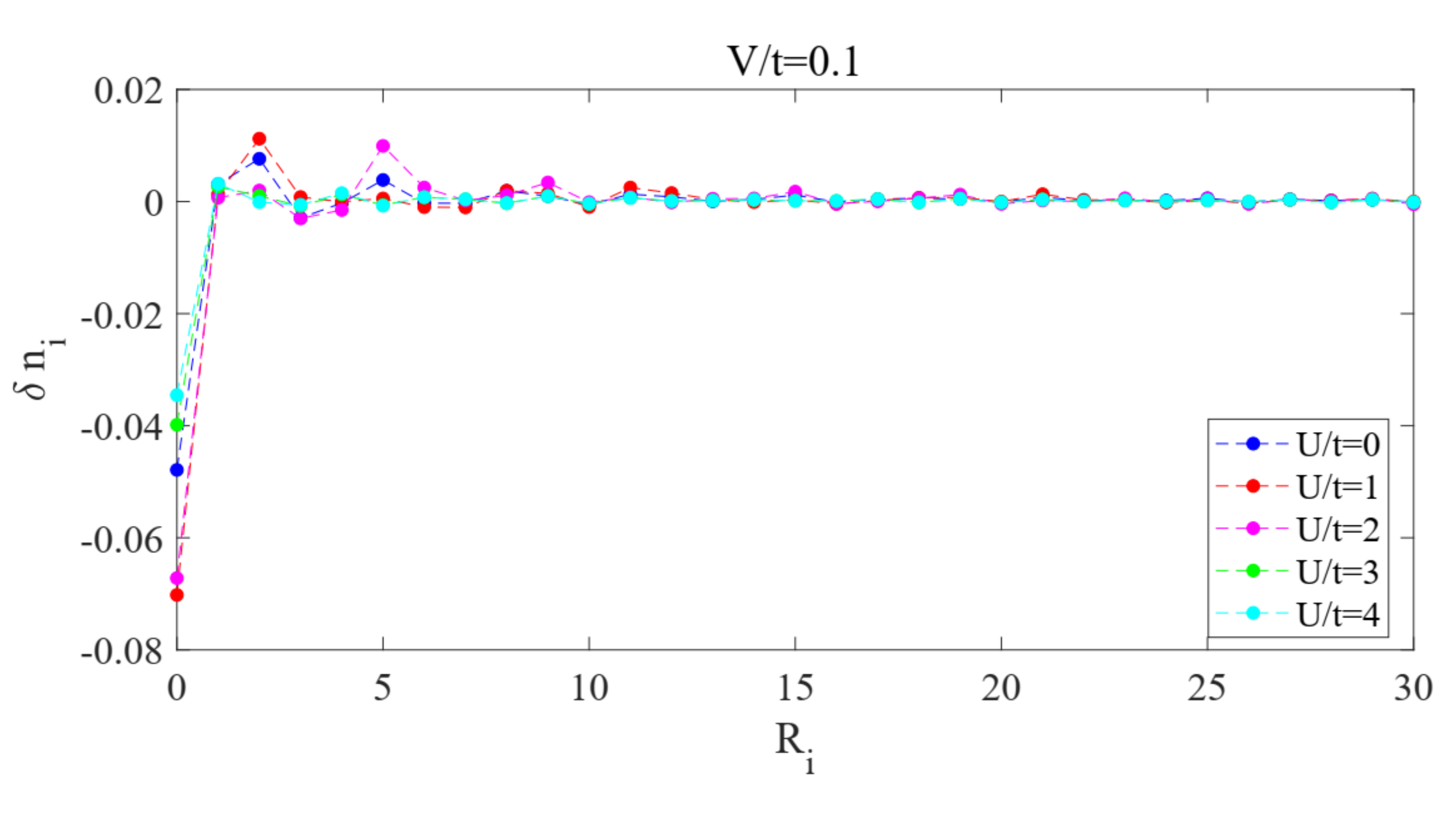}
\caption{\label{fig:V_01_LRT_06} $\delta n_{i}$ for $U/t=0,1,2,3,4$ with $V/t=0.1$ and electron density is fixed to $0.6$.}
\end{figure}
\section{$\delta n_{i}$ for electron density $n=0.4$ and $n=0.6$ from linear response theory}\label{ap_C}

In the main text, we have given the FO results for electron density $n=0.5$ in terms of linear response theory. To see whether the physics depends on the choice of electron density, here the results for $n=0.4$ and $n=0.6$ are shown in Figs.~\ref{fig:V_01_LRT_04} and \ref{fig:V_01_LRT_06}. It is clearly that, just as the case of $n=0.5$, there also exists FO for $n=0.4$ and $n=0.6$. Therefore, we may say that the NFL state indeed exhibits FO for generic electron filling.

\end{document}